\documentclass[fleqn,usenatbib]{mnras}

\usepackage{newtxtext,newtxmath}

\usepackage{hyperref}
\usepackage[T1]{fontenc}
\usepackage{ae,aecompl}
\usepackage{adjustbox}
\usepackage{longtable}
\usepackage[english]{babel}

\usepackage{graphicx}	
\usepackage{amsmath}	
\usepackage{amssymb}	

\title{Multiple AGN activity during the BCG assembly of XDCPJ0044.0-2033 at z$\sim$1.6}
        
\author[A. Travascio et al.]{
A. Travascio,$^{1}$\thanks{E-mail: travascio.andrea91@gmail.com}
A. Bongiorno,$^{1}$
P. Tozzi,$^{2}$
R. Fassbender,$^{3}$
F. De Gasperin,$^{4}$
V. F. Cardone,$^{1,5}$\newauthor
L. Zappacosta,$^{1}$
G. Vietri,$^{6}$
E. Merlin,$^{1}$
M. Bischetti,$^{1}$
E. Piconcelli,$^{1}$
F. Duras,$^{7}$\newauthor
F. Fiore,$^{8}$
N. Menci,$^{1}$
P. Mazzotta,$^{9}$
and A. Nastasi$^{10}$
\\
$^{1}$INAF - Osservatorio Astronomico di Roma, Via di Frascati 33, 00040, Monteporzio Catone, Rome, Italy\\
$^{2}$INAF - Osservatorio Astrofisico di Arcetri - Largo E. Fermi 5 I-50125\\
$^{3}$Max-Planck-Institut f\"ur extraterrestrische Physik, Giessenbachstrasse 1,85748 Garching, Germany\\
$^{4}$Hamburger Sternwarte, Universit\"at Hamburg, Gojenbergsweg 112, 21029, Hamburg, Germany\\
$^{5}$Istituto Nazionale di Fisica Nucleare, Sezione di Roma 1 - Piazzale Aldo Moro 00185, Rome, Italy\\
$^{6}$INAF - Istituto di Astrofisica Spaziale e Fisica Cosmica - Milano, via A. Corti 12, I-20133, Milano, Italy\\
$^{7}$Aix Marseille Univ, CNRS, CNES, LAM, Marseille, France\\
$^{8}$INAF - Osservatorio Astronomico di Trieste - Via G. Tiepolo 11, Trieste, Italy\\
$^{9}$Dipartimento di Fisica, Universita' di Roma "Tor Vergata" Via della Ricerca Scientifica 1, I-00133 Rome, Italy\\
$^{10}$GAL Hassin - Centro Internazionale per le Scienze Astronomiche, Isnello, Italy\\
}

\date{August 25, 2020}

\pubyear{2019}
\hypersetup{draft}
\begin{document}
\label{firstpage}
\pagerange{\pageref{firstpage}--\pageref{lastpage}}
\maketitle

\begin{abstract}
Undisturbed galaxy clusters are characterized by a massive and large elliptical galaxy at their center, i.e. the Brightest Cluster Galaxy (BCG). How these central galaxies form is still debated. According to most models, a typical epoch for their assembly is z$\sim$1-2.
We have performed a detailed multi-wavelength analysis of the core of XDCP J0044.0-2033 (\texttt{XDCP0044}), one of the most massive and densest galaxy clusters currently known at redshift z$\sim$1.6, whose central galaxy population shows high star formation compared to lower-z clusters and an X-ray AGN located close to its center.
SINFONI J-, H- and KMOS YJ-, H- bands spectroscopic data have been analyzed, together with deep archival HST photometric data in F105W, F140W, and F160W bands, Chandra X-ray, radio JVLA data at 1-2 GHz, and ALMA band-6 observations.
In the very central region of the cluster ($\sim$70~kpc $\times$ 70~kpc), two systems of interacting galaxies have been identified and studied (\texttt{Complex A} and \texttt{B}), with a total of seven confirmed cluster members. 
These galaxies show perturbed morphologies and three of them show signs of AGN activity. In particular, two type-1 AGN  with typical broad lines have been found at the center of each complex (both of them X-ray obscured and highly accreting with $\rm \lambda_{Edd}\sim 0.4-0.6$), while a type-2 AGN has been discovered in Complex A. The AGN at the center of \texttt{Complex B} is also detected in X-ray while the other two are spatially related to radio emission. The three AGN provide one of the closest AGN triple at z$>$1 revealed so far with a minimum (maximum) projected distance of 10 (40) kpc.  
The observation of high star formation, merger signatures and nuclear activity in the core of \texttt{XDCP0044} suggests that all these processes are key ingredients in shaping the nascent BCG. According to our data, \texttt{XDCP0044} could form a typical massive galaxy of $\rm M_{\star}\sim 10^{12} M_{\odot}$, hosting a Black Hole (BH) of $\rm 2 \times 10^8 - 10^9 M_{\odot}$, in a time scale of the order of $\sim$2.5 Gyrs.
\end{abstract}




\section{Introduction}

Brightest cluster galaxies (BCGs) are the most massive ($\rm M_{\star} \sim 10^{12} M_{\odot}$) and luminous ($\rm M_v \approx -23$) galaxies and reside at the center of relaxed, virialized and undisturbed galaxy clusters in the local Universe (\citealt{Sandage76,Lin10}). BCGs are usually located at the minimum of the cluster potential well, close to the peak of the X-ray emission \citep{Jones84}. At low redshift, they appear like red, quiescent, massive and large \citep[][up to 100 kpc]{Carter77,Bernardi07} elliptical galaxies \citep{Dubinski98} and they often show radio nuclear activity and jet emissions, able to affect the gas in the Intra Cluster Medium \citep[ICM,][]{Best07,Hogan15,Moravec20}. On the contrary, most observations of galaxy cluster cores at z>1 do not show the presence of a single BCG but are instead characterized by star-forming galaxies (SFGs) with disturbed morphology \citep{Zhao17}. 
Moreover, although in some cases galaxies exhibiting early type morphology are also found in z$\geq$1.4 cluster cores \citep{Strazzullo13,Cooke16}, evidence is reported by several authors for an increase of the SFGs fraction in these distant environments \citep{Bai07,Bai09,Krick09}. Specifically, in high-z systems, a reversal star formation (SF) - density relation has been observed, i.e. while at $z<1.4$ the number of SFGs increases towards the cluster outskirts, at $z>1.4$ the SFGs fraction is higher in the cluster cores \citep{Brodwin13,Santos15}. 
To date, the process responsible for the transition between the unquenched and quenched eras in such environments is still not clear and its understanding is fundamental not only for explaining the formation of local BCGs but also for both cluster and galaxy evolution theory \citep{Lin04,Rudick11,Contini14}.

Because the BCGs are located in privileged places in the cluster cores, these are expected to experience SF processes, AGN feedback and multiple mergers. 
According to many cosmological simulations and semi-analytic models \citep[e.g.][]{Springel05,Croton06,DeLucia07,Cooke16,LeeBrown17,RagoneFigueroa18,Pillepich18} the BCG progenitors form most ($\sim$50\%) of their stars at z$>$2.5. This mass is then assembled at z$\sim$1-2 through mergers and, later (z<1), through multiple accretion of small galaxies, to form the final BCG \citep{Stott08,Lidman13,Laporte13}.
On the other hand, recent IR and sub-mm observations of molecular cold gas in galaxy clusters at 1$<$z$<$2, found a significant SFR (tens and hundreds $\rm M_{\odot}~yr^{-1}$) in BCG \citep{Webb15b,McDonald16} and this is mainly attributed to wet major and minor mergers \citep{Webb17,Bonaventura17}. According to these works, z$\sim$1-2 should be a crucial epoch during which we expect a high rate of merger activity among galaxies (the BCG progenitors) residing in the core of massive galaxy clusters. These mergers will eventually lead to deposit large reservoir of gas in cluster cores and induce dust-obscured starburst events \citep{Cooke19}.

Most of the studies found indeed an enhancement of the merger activity in high-z galaxy clusters with respect to the low redshift systems and the field ones \citep{Fassbender11,Mancone12,Lotz13,Lidman13,Mei15}. 
In particular, \citet{Alberts16} found an excess of SF and merger activity in the cores of massive ($\rm M > 10^{14} M_{\odot}$) galaxy clusters at z>1.
Conversely, other works found the merger fraction in galaxy clusters \citep{Andreon13} and/or their cores \citep{Delahaye17} comparable to the field one.
Interestingly, \citet{Alberts16} also found an excess of AGN activity in such high-z systems, highlighting the key role of dense environments in triggering nuclear activity at this epoch and pointing towards a co-evolution between SF and Black Hole (BH) accretion \citep{Galametz09,Martini13}. 
Moreover, the excess of both SF and AGN activity in high-z galaxy clusters relative to the field one, supports a scenario in which (i) the increase in the merging rate at z>1 is responsible for triggering both formation of stars and BH accretion, and (ii) the subsequent possible AGN feedback could be one of the mechanisms able to quench SF in massive galaxies, leading to the formation of red and dead elliptical galaxies, as observed in local galaxy cluster cores \citep{Springel05,Hopkins08,Narayanan10}. 
This would also explain the inversion of the SF-density relation observed in z>1.4 clusters \citep[see Fig.~6 in][]{Brodwin13}. In fact, while the higher normalisation of the relation at higher redshift is easily explained  with the higher number of mergers, the change in the slope implies a mechanism able to stop SF (e.g. AGN feedback) in shorter time scales in galaxies located in the cluster core than in the outer region. 

\begin{figure*}
\centering
\includegraphics[width=7.2in]{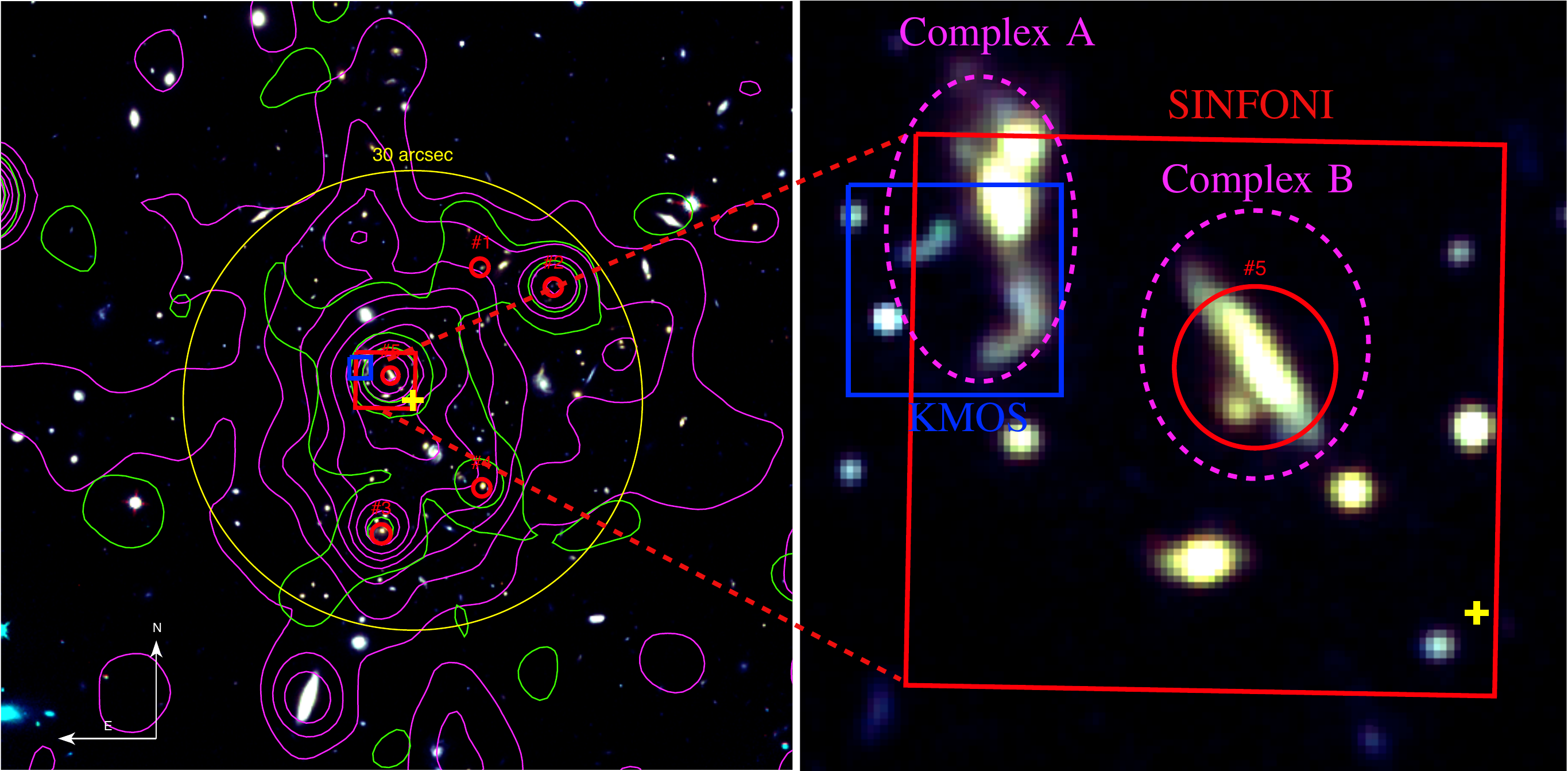}
\caption{\textit{Left Panel:}: HST RGB (F105W + F140W + F160W bands) image of the galaxy cluster \texttt{XDCP0044}. The yellow cross indicates the centroid of the extended X-ray emission (RA 00:44:05.2, Dec -20:33:59.8) while in yellow is reported a circle of radius $\sim 30''$, corresponding to 250 kpc. The green and magenta contours are the Chandra hard ($\rm [2-7] keV$) and soft ($\rm [0.5-2] keV$) X-ray emissions, respectively, while the red circles mark the unresolved X-ray sources as identified by \citet{Tozzi15}. Finally, the red and blue squares delimit the region analyzed in this paper, corresponding to the SINFONI and the KMOS field of views. \textit{Right Panel}: zoom-in of the analyzed central region. The green circle encloses the region covered by the ALMA observations. The magenta dashed lines mark the two detected complex of galaxies. Specifically, the complex in the upper left corner of the SINFONI field (\texttt{Complex A}) is the BCG as identified by \citet{Fassbender14} while the central complex (\texttt{Complex B}) is the X-ray AGN identified by \citet{Tozzi15} (red circle). }
\centering
\label{HSTzoom}
\end{figure*}

In this paper we studied the core of the X-ray detected galaxy cluster XDCP J0044.0-2033 (hereafter \texttt{XDCP0044}; \citealt{Santos11,Fassbender11}) at z$\sim$1.6.
\texttt{XDCP0044} is one of the most massive galaxy clusters discovered at $\rm z>1.4$ with $\rm M_{200}$\footnote{The mass which encloses an over-density of 200 times the critical density of the Universe, computed by  \citet{Tozzi15} by modelling the X-ray data.}$\rm \approx 4.4 _{-0.8} ^{+1.3} \times 10^{14} M_{\odot}$. It has been detected in low-resolution XMM archival data within the XMM-Newton Distant Cluster Project (XDCP, \citealt{Fassbender11}) thanks to its extended X-ray emission (RA=00:44:05.2, Dec = -20:33:59.7), and confirmed by deep Chandra observation to have strong diffuse emission typical of virialized clusters.
\texttt{XDCP0044} is in a quite advanced state of dynamical relaxation, with evidence of ongoing cluster-scale major-merger activity \citep{Fassbender14,Tozzi15} and of a reversal star formation-density relation \citep{Santos15}.  
Fig.~\ref{HSTzoom} (left panel) shows the HST RGB (as a combination of F105W, F140W and F160W filters) image of  \texttt{XDCP0044} with overlaid the X-ray Chandra soft ([0.5-2] keV, magenta) and hard ([2-7] keV, green) bands contours. Red circles indicate the 5 point-like sources (AGN) identified by \citet{Tozzi15} within 30$''$ ($\sim$250 kpc) from the cluster center. 
\texttt{XDCP0044} is a unique laboratory to study the building-up of the BCG and the interplay between galaxies, nuclear activity, and the inter-galactic gas in the core of massive high redshift galaxy clusters.
In this work, we have conducted a detailed multi-wavelength study of the very central region ($\sim$70 kpc$~\times~$70 kpc) of the \texttt{XDCP0044} core (right panel of Fig.~\ref{HSTzoom}). 

The paper is organised as follows: Sect.~\ref{sec:ObsData} describes the observations and data reduction, while Sect.~\ref{sec:analysis} presents the photometric and spectroscopic analysis of the galaxies in the analyzed region. We then focus on the AGN and SF activity in Sect.~\ref{sec:AGNanalysis}. 
Discussion, Summary and Conclusions are presented in Sect.~\ref{sec:discussion} and \ref{sec:conclusion}.

Throughout the paper we will assume a cosmology with $\Omega_\Lambda=0.6842$ and $\rm H_0=67.32 \,  km\, s^{-1} Mpc^{-1}$ \citep{Plank18}, and the errors will be quoted at $1\sigma$ and upper/lower limits at $90\%$ confidence level, unless otherwise stated.

\section{Observations and Data Reduction} \label{sec:ObsData}
In this paper we combine and study the information derived from several photometric and spectroscopic multi-wavelength observations of the inner region of \texttt{XDCP0044}, from X-ray to optical, near-infrared (NIR) and radio bands, that have been obtained in the recent years. 
A summary of the analysed data is presented in this section.

\subsection{Proprietary data}

\subsubsection{SINFONI observations and data reduction}\label{sinfodat}
Integral Field Unit (IFU) SINFONI observations in J- and H-band of the central region of \texttt{XDCP0044} (program ID 094.A-0713(A); PI A. Bongiorno), have been obtained in 2014. At the redshift of the cluster, J-band corresponds to the rest-frame [OIII] and H$\beta$ emission lines ($\sim$ 4496 \AA\ - 5230 \AA) while H-band samples the H$\alpha$ emission line with a resolution of R=2000 and R=3000, respectively.\\
The data have been taken in seeing limited mode (average seeing $\sim$ $0.8''$) in a $8'' \times 8''$ field of view (FOV) which corresponds to $\sim$70 kpc $\times$ 70 kpc. The observed FOV (centered at RA = 00:44:05.420, DEC = $-$20:33:57.16) is shown in the right panel of Fig.~\ref{HSTzoom} with a red square.
Observations consist of 6 Observing Blocks (OBs) in J-band and  2 OBs in H-band. We performed observations of 300 s per frame both on objects (O) and on sky (S), following the scheme "OOSSOOSSOO", in addition to the observations of the standard and telluric stars. The total on target integration time is $\sim$4 h in J-band and 1.5 h in H-band. 
SINFONI data reduction was performed using the ESO pipeline \texttt{ESOREX} \citep{PipelineEsorex}, with the improved sky subtraction proposed by \citet{Davies2007}. After flat-fielding, dark correction, correction for distortions, cosmic rays removal and wavelength calibration, each frame within a single OB was corrected for the sky emission lines, using the IDL routine \texttt{skysub.pro} \citep{Davies2007}. 
The science frames within each OB were combined considering the offsets of the object in each frame and, finally, flux calibrated according to the standard stars. Moreover, a further flux correction in J-band was applied by rescaling the continuum flux to the HAWK-I J-band photometric point published in \citet{Fassbender14}. 
The flux calibrated exposures of the different OBs were then combined together by measuring and applying the relative offset between the peak emissions of the most luminous sources in the fields. The final result of the data reduction procedure was a 3D flux-calibrated data cube having a PSF FWHM$\sim$0.7$''$ and 0.6$''$ in J- and in H-band, respectively.

\subsubsection{KMOS observations and data reduction}

KMOS IFU observations in JY- and H-bands have been obtained in 2013 (Program ID: 092.A-0114(A); PI R. Fassbender). 
In this work we focused on the analysis of the KMOS data centered on the BCG candidate identified by \citet{Fassbender14}. 
At the redshift of the cluster, JY-band samples the [OIII] and H$\beta$ emission lines with a resolving power $\rm R\simeq$3600 while H-band samples the H$\alpha$ region with $\rm R\simeq$4000.
The data have been taken with an average seeing of $\sim$ $1''$  in a $2.8'' \times 2.8''$ FOV. The observed FOV, centered at RA = 00:44:05.600, DEC = -20:33:54.716, is shown in Fig.~\ref{HSTzoom} with a blue square. 
Observations consist of 7 OBs in YJ-band and 2 OBs in H-band. Each OB is made of 5 frames, each of 450 s integration time, for both sky (S) and science observations (O). The total on target integration time is $\sim$4.4 h in YJ-band and $\sim$1.25 h in H-band.

The data were reduced using the pipeline with the Software Package for Astronomical Reduction with KMOS (SPARK; \citealt{Davies13}), which includes dark correction, flat fielding, illumination correction, wavelength calibration and the sky subtraction \citep{Davies11}. Data were then combined according to the spatial shift of the object in each frame. The final datacubes in JY- and H- bands have a PSF FWHM of $\sim$0.8$''$ and 0.9$''$, respectively.

\subsubsection{JVLA observations and data reduction}

JVLA data have been taken in 2016 in the L-band (1--2 GHz) in A, B, and C configuration (with $\rm t_{exp}$ of 5 h each) as part of the project 16A-082 (PI F. De Gasperin). The data were reduced using the \texttt{CASA}\footnote{\url{https://casa.nrao.edu/}} package. The visibilities were Hanning-smoothed, bandpass-calibrated and then flagged using the automatic tool \texttt{AOflagger} \citep{Offringa12}. We used 3C\,147 as flux calibrator and the point source J2357--1125 as phase calibrator. The flux scale has been set to \citet{Perley13}. Bandpass, scalar delays, cross-hand delays, and polarisation angle corrections were transferred to the target and the phase calibrator. Then, phase and rescaled amplitude from the phase calibrator were transferred to the target field. Finally, a few cycles of phase-only self-calibration was applied on the target field. We obtained final images in A, B and C configurations with resolution of $2'' \times 1''$, $6'' \times 3''$ and $16'' \times 8''$ and noise of 15, 20 and 25 $\rm \mu Jy/beam$, respectively. The datasets were finally combined to obtain single image used for the scientific analysis. 

We note that all flux density errors for extended emission were computed as $S_{\rm err} = \sigma \cdot \sqrt{N_{\rm beam}}$, where $\sigma$ is the local image rms and $N_{\rm beam}$ is the number of beams covering the source extension.

\subsection{Archival data}\label{sec:hstdata}

\textbf{HST} data of the \texttt{XDCP0044} galaxy cluster have been obtained in 2015 (Program 13677; PI S. Perlmutter) in F105W, F140W, F160W and F814W bands, with the following exposure times: 2 h in F105W and F140W, 1 h in F160W and 47 minutes in F814W. In our analysis we used the images obtained by combining the archival $drizzled$ (DRZ) frames, after performing the astrometry and aligning them. Due to the low S/N, the F814W band has not been included in the analysis.

\textbf{Chandra/ACIS-S} X-ray observation of the cluster has been performed in October-December 2013 and carried out in six exposures for an total time of $\sim$370 ks (PI P. Tozzi). Details on observations and data reduction can be found in  \citet{Tozzi15}. In Sect.~\ref{sec:xray} we present a detailed spectral analysis.

\textbf{ALMA} band 6 data from project 2017.1.01387.S (PI S. Stach), covering the FOV of our SINFONI data, are also included in our study. These observations span the frequency range 221.5-225.3 GHz and 236.3-240.7 GHz for a total bandwidth of $\sim$7.5 GHz, and a spectral resolution of $\rm \sim 40~km~s^{-1}$. Given the redshift of the cluster, the ALMA coverage corresponds to the rest-frame $\sim$600~GHz continuum and CO(5-4) emission line ($\rm \nu_{rest}$=576.27~GHz).  ALMA data were calibrated with the CASA 5.1.1 version in pipeline mode using the default calibrators provided by the Observatory. The continuum map was created by averaging the visibilities over the whole bandwidth. Specifically, we applied the standard non-interactive cleaning, with clark algorithm and natural weighting. This resulted into a rms sensitivity of 0.024~mJy/beam and an angular resolution of 0.98 $\times$ 0.78 $\rm arcsec^2$.
The resulting sensitivity to the native spectral resolution of the observations ($\sim$10$\rm~km~s^{-1}$) is 1.5 mJy/beam and the beam is 0.82 $\times$ 0.63 arcsec$^2$.

\section{Analysis}\label{sec:analysis}

\subsection{Optical/NIR source identification} \label{sec:Photometry}

\begin{figure}
\includegraphics[width=0.48\textwidth]{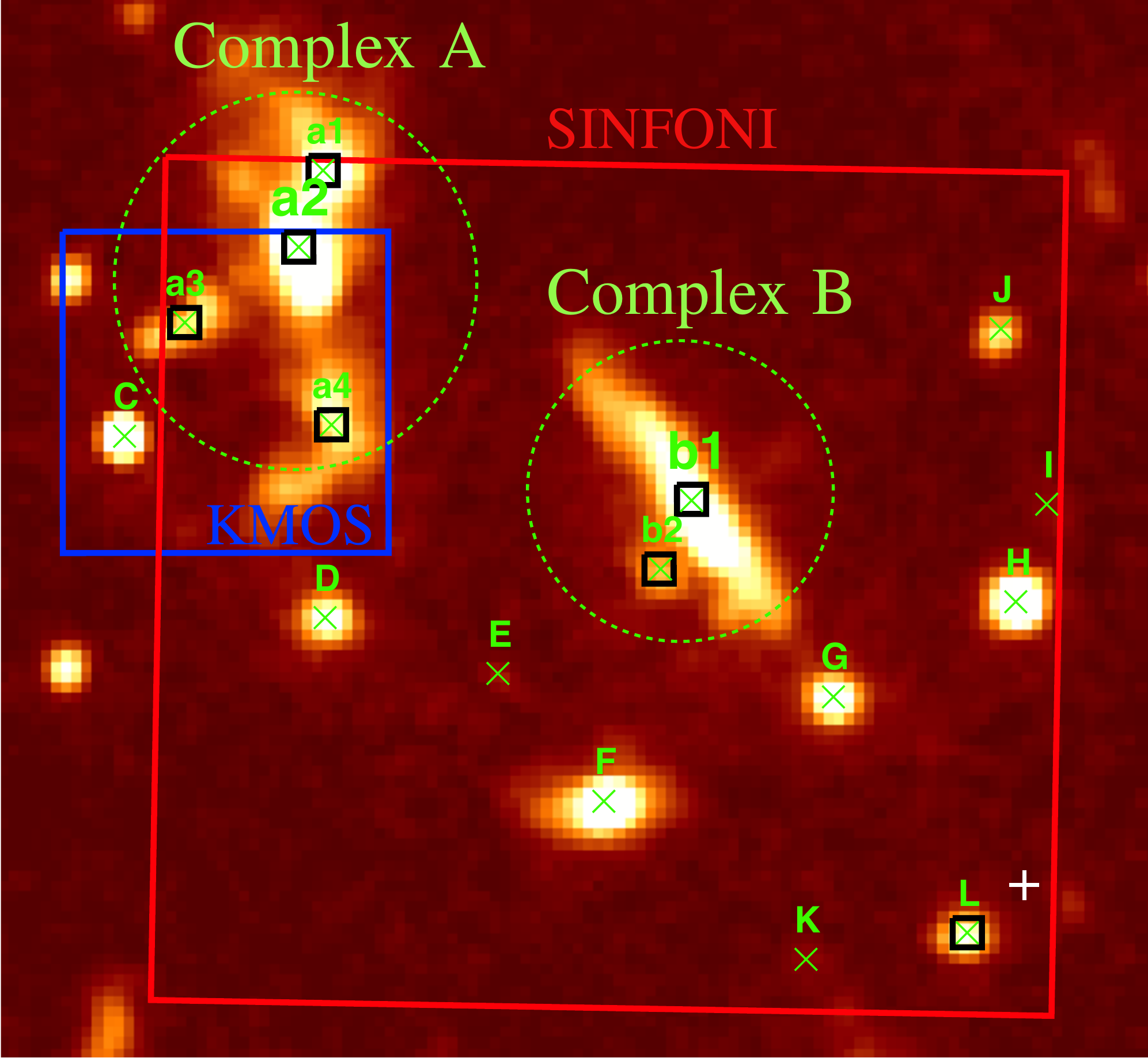}
\centering
\caption{HST F105W image of the core of \texttt{XDCP0044}, where two complexes (A and B) are highlighted with the green dashed line. Green crosses mark the sources identified in the HST images and listed in Table \ref{tab:FluxHST}. Black squares mark the sources for which a redshift has been measured while the white cross is the X-ray centroid.}
\centering
\label{fig:HST4}
\end{figure}

\begin{table}
\caption{HST-band  magnitudes in AB System of the identified sources.}
\label{tab:FluxHST}
\bigskip
\centering\small\setlength\tabcolsep{3pt}
\begin{adjustbox}{width=\columnwidth,center}
\begin{tabular}{c|c|c|c|c|c}
$ID$  & RA   &  DEC & $m_{F105W}$ & $m_{F140W}$ & $m_{F160W}$  \\ [2pt]
\hline 
a1 & 00:44:05.595 & -20:33:53.61 &   $22.756  \pm  0.140$ &   $22.087  \pm  0.097$ &  $21.858  \pm  0.162$  \\
a2 & 00:44:05.607 & -20:33:54.51 &  $22.288  \pm  0.113$ &   $21.600  \pm  0.077$ &  $21.363  \pm  0.129$     \\
a3 & 00:44:05.680 & -20:33:54.92 &  $23.602  \pm  0.206$ &   $23.217  \pm  0.163$ &  $23.214  \pm  0.303$     \\
a4 & 00:44:05.590 & -20:33:55.80 &  $22.465  \pm  0.123$ &   $21.962  \pm  0.092$ &  $21.776  \pm  0.156$  \\
\hline
b1  & 00:44:05.369 & -20:33:56.45 &  $21.717  \pm  0.086$ &   $21.014  \pm  0.059$ &  $20.815  \pm  0.100$         		   \\
b2    & 00:44:05.388 & -20:33:57.04 &  $25.225  \pm  0.435$ &   $24.340  \pm  0.274$ &  $24.058  \pm  0.446$   \\
\hline
C    & 00:44:05.717 & -20:33:55.90  &  $23.277  \pm  0.178$ & $23.123 \pm 0.156$ &  $23.032  \pm  0.279$    \\
D    & 00:44:05.594 & -20:33:57.46 &  $23.357  \pm  0.186$ &   $22.797  \pm  0.135$ &  $22.620  \pm  0.231$      \\
E    & 00:44:05.488 & -20:33:57.94 &  $25.974  \pm  0.647$ &   $25.441  \pm  0.479$ &  $25.295  \pm  0.824$ \\
F    & 00:44:05.423 & -20:33:59.04 &  $22.466  \pm  0.123$ &   $21.569  \pm  0.077$ &  $21.367  \pm  0.129$  \\
G     & 00:44:05.282 & -20:33:58.14 &  $23.368  \pm  0.188$ &   $22.492  \pm  0.118$ &  $22.290  \pm  0.198$  \\
H    & 00:44:05.170 & -20:33:57.32 &  $23.049  \pm  0.161$ &   $22.436  \pm  0.115$ &  $22.248  \pm  0.195$    \\
I    & 00:44:05.151 & -20:33:56.48 &  $27.047  \pm  1.028$ &   $27.042  \pm  0.980$ &  $26.666  \pm  1.502$ \\
J    & 00:44:05.179 & -20:33:54.97 &  $24.162  \pm  0.274$ &   $23.763  \pm  0.217$ &  $23.514  \pm  0.354$   \\
K     & 00:44:05.299 & -20:34:00.40&  $26.192  \pm  0.702$ &   $25.833  \pm  0.566$ &  $25.490  \pm  0.883$   \\
L    & 00:44:05.200 & -20:34:00.17 &  $24.106  \pm  0.264$ &   $23.684  \pm  0.205$ &  $23.554  \pm  0.357$  		\\
\hline
\end{tabular}
\end{adjustbox}
\end{table}

We limited our analysis of HST data to the central region of the cluster, where spectroscopic SINFONI and KMOS observations have been taken (right panel of Fig.~\ref{HSTzoom}) and we used \textsc{SExtractor} \citep{Bertin96} to detect and deblend the sources. We chose F105W as detection band, identifying a total of 16 sources (see Fig.~\ref{fig:HST4}).  
Interestingly, two main galaxies complexes have been identified: 
\begin{itemize}

\item \texttt{Complex A}, in the top-left corner of the \texttt{XDCP0044} core (Fig.~\ref{fig:HST4}), consists of several sources identified in HST data. This complex includes what was identified by \citet{Fassbender14} in the HAWK-I images as a single source and classified as the BCG, although with several extensions, interpreted as sign of ongoing or recent mergers.\\

\item{} \texttt{Complex B} in the central region of the SINFONI field. This complex includes the X-ray AGN discovered by \cite{Tozzi15} (source b1 in Fig.~\ref{fig:HST4}). The HST photometric analysis revealed the presence of a second source (\texttt{b2}) very close to the central one.

\end{itemize}

\begin{figure*}
	\centering
	\includegraphics[width=1.0\textwidth]{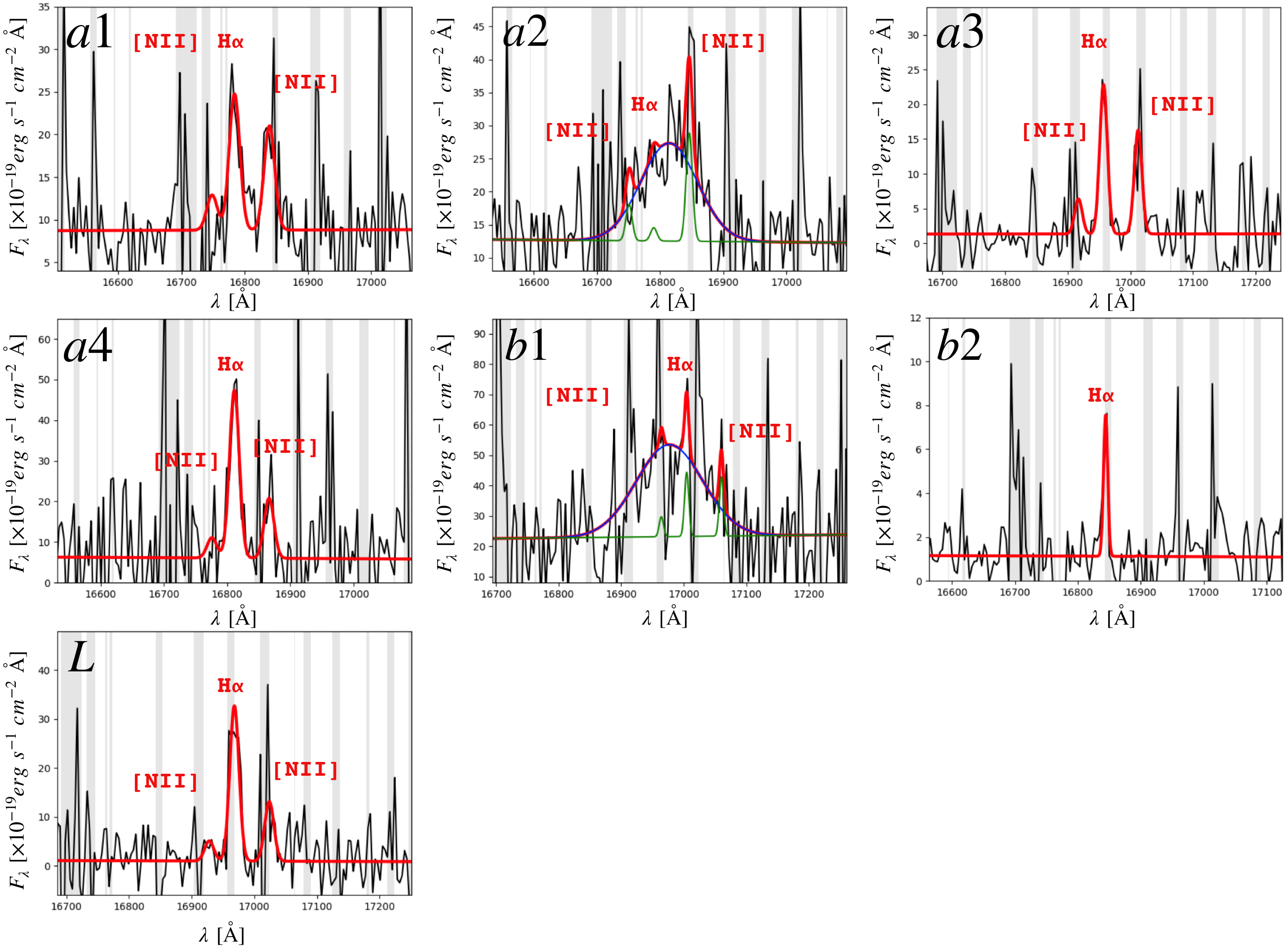}
     \caption{Zoom-in of the spectra (smoothed at $\sim$70 $\rm km~s^{-1}$) around the H$\alpha$ line of the 7 confirmed cluster members. The red line represents the total best-fit model consisting of a power-law for the continuum plus up to 4 Gaussian components to model the emission lines. The fit to the continuum and the narrow lines are shown in green while the broad components in blue.}
	\label{fig:spectra2}
\end{figure*}

For the 16 sources, aperture photometry was performed in \textit{dual-mode} on the available bands. We did not PSF-match the images, as the FWHMs in all images are comparable ($\sim$ 0.130", 0.137",0.145" in F105W, F140W and F160W bands, respectively).
We took the isophotal flux\footnote{\textsc{SExtractor} \texttt{FLUX\_ISO}, namely the summation of the fluxes in all the pixel assigned to each object in the segmentation map.} as the best photometric estimate, since the severe blending of sources would make photometric measurements in larger areas unreliable. In particular, we do not attempt to estimate total fluxes using Kron apertures \footnote{Elliptical aperture defined by the second order moments of the object's light distribution in \textsc{SExtractor} routine.} (\texttt{FLUX\_AUTO}), as they would be strongly contaminated by light coming from neighbouring sources. 
AB system magnitudes (listed in Table \ref{tab:FluxHST}) have been estimated from the isophotal flux using zero points zp=(26.2, 26.4, 25.9) for (F105W, F140W, F160) band, respectively.

\subsection{NIR spectroscopy} \label{sec:spectroscopic}

\begin{table*}
\begin{center}    
\caption{Properties derived from the spectral analysis. Columns are: object ID, redshifts, FWHM and flux of the H$\alpha$ line, luminosity at 5100\AA~and V-band absolute magnitude $\rm M_V$. Upper limits are given at 3$\sigma$.}             
\label{tab:7emsources}     
\begin{adjustbox}{width=\textwidth}
\begin{tabular}{c c c c c c c c c c}         
\hline\hline                       
$\rm ID$ & $z$   & $\rm FWHM$ (H$\alpha$) & $\rm F(H \alpha)$ & $\rm F([NII] \lambda 6549 \AA)$ & $\rm F([NII] \lambda 6585 \AA)$ & $\rm F([OIII] \lambda 5070 \AA)$ & $\rm F(H \beta)$ & $\rm log(L_{5100})$ & $\rm M_V$ \\
&&   $\rm [km\,s^{-1}]$  &   $\rm \times 10^{-17}[erg\,s^{-1}~cm^{-2}]$   &  $\rm \times 10^{-18} [erg\,s^{-1}~cm^{-2}]$ & $\rm \times 10^{-18} [erg\,s^{-1}~cm^{-2}]$ &   $\rm \times 10^{-17}[erg\,s^{-1}~cm^{-2}]$ &   $\rm \times 10^{-18}[erg\,s^{-1}~cm^{-2}]$ & $\rm erg~s^{-1}$ &  \\
\hline
a1   & 1.5567 &  339 $\pm$ 28 & 3.23 $\pm$ 0.40 & 8.2 $\pm$ 1.0  & 24.7 $\pm$ 3.1 & <0.04 & <0.27 & $44.25 \pm 0.21$ & -23.1  \\
a2   & 1.5577 & 1880 $\pm$ 321 $^b$ & 16.62 $\pm$ 4.36 $^b$ & 6.1 $\pm$ 1.1 & 18.0 $\pm$ 3.2 & $1.96 \pm 0.30$ & $10.8 \pm 1.65$ & $44.45 \pm 0.17$ & -23.6    \\
a3   & 1.5831 &  175 $\pm$ 82 & 3.13 $\pm$ 1.53 & 7.3 $\pm$ 3.6  & 21.7 $\pm$ 10.6 & $1.83 \pm 0.50$ & $3.66 \pm 0.97$ & $43.83 \pm 0.36$ & -22.1   \\
a4   & 1.5609 &  311 $\pm$ 108 & 7.67 $\pm$ 4.00 & 9.3 $\pm$ 4.8  & 27.5 $\pm$ 14.3 & <0.11 & <0.48 & $44.31 \pm 0.20$ & -23.2   \\
b1   & 1.5904 & 2205 $\pm$ 383 $^b$ & 40.23 $\pm$ 10.72  $^b$ & 4.0 $\pm$ 1.4  & 11.7 $\pm$ 4.1 & <0.03 & <0.38 & $44.71 \pm 0.13$ & -24.2    \\
b2   & 1.5659 &  100 $\pm$ 62 & 0.39 $\pm$ 0.31 & <0.70 & <1.95 & <0.07 & <0.44 & $43.35 \pm 0.71$ & -20.8        \\
L    & 1.5848 &  290 $\pm$ 99 & 5.55 $\pm$ 2.80 & 7.2 $\pm$ 3.7 & 21.2 $\pm$ 10.8 & <0.15 & <0.67 & $43.65 \pm 0.47$ & -21.6  \\
\hline
\end{tabular}
\end{adjustbox}
\end{center}
Note: FWHM reported in table have been corrected for the instrumental ; $^b$ refers to the broad H$\alpha$ component. 
\end{table*}

In this section, we present the spectroscopic analysis of VLT/SINFONI and KMOS IFU data. 

A first extraction of the spectra of the sources identified in HST was performed using a fixed aperture diameter of 7 pixels in SINFONI (0.875$''$) and 5 pixels in KMOS (1$''$) and each spectrum was normalized using the HST photometry.
For 7 out of the 16 galaxies, a clear H$\alpha\lambda6563\text{\AA}$ line, one of the strongest line expected, has been identified together with few other lines (e.g. [OIII]$\lambda\lambda 4959\text{\AA}5007\text{\AA}$ doublet, H$\beta\lambda4861\text{\AA}$, [OII]$\lambda3727\text{\AA}$ and [NII]$\lambda\lambda6550\text{\AA},6585\text{\AA}$), which confirmed the redshift.
The spectra are shown in Figs.~\ref{fig:spectraTot1} and \ref{fig:spectraTot2} and their redshifts, listed in table~\ref{tab:7emsources}, range from z=1.5567 to z=1.5904  ($\Delta z \simeq 0.0337$), consistently with the redshift of the cluster. 
All 7 galaxies (shown with a black square in Fig.~\ref{fig:HST4}) are therefore spectroscopically confirmed cluster members. 
Four of them (\texttt{a1}, \texttt{a2}, \texttt{a3} and \texttt{a4}) belong to \texttt{Complex A}, other two are instead part of \texttt{Complex B}, and the \texttt{L} is close to the X-ray centroid. The spectroscopic analysis confirms that \texttt{Complex A} and \texttt{B} are indeed interacting galaxies of at least 4 galaxies at a projected distance of 20 kpc in \texttt{Complex A} and 2 galaxies at $\sim$ 5~kpc in \texttt{Complex B}. The two complexes are very close to each other, i.e. $\sim~35~$kpc.
Moreover, as detailed later, in two of the analyzed sources, \texttt{a2} in \texttt{Complex A} and \texttt{b1} in \texttt{Complex B}, the H$\alpha$ emission line is broad ($\rm FWHM> 1800~\rm{km~s^{-1}}$). These sources have been therefore classified as broad line AGN (BL-AGN).

Given the low quality of the data, for these 7 sources, we built the integrated Signal to Noise Ratio (SNR) map of the H$\alpha$ emission lines and we extracted the spectra by using a region centered on the H$\alpha$ line SNR peak, with a radius of $\sim$0.4$''$. Fig.~\ref{fig:spectra2} reports a zoom-in of the H$\alpha$ spectral region together with the best-fit to the data (red line) consisting of a power law for the continuum, plus up to 4 Gaussian components which model the narrow and broad H$\alpha$ line and the doublet [NII]$\lambda\lambda6550\text{\AA},6585\text{\AA}$ emission lines. In the fit, we constrained the intensity of [NII]$\lambda 6585\text{\AA}$ to be 2.96 times the [NII]$\lambda 6550\text{\AA}$ one \citep{Acker89} and we set all three narrow lines to have the same dispersion.
The results of the best-fit model, i.e FWHM corrected for the instrumental resolution, H$\alpha$ and [NII] (if detected) fluxes, are reported in Table~\ref{tab:7emsources}. Uncertainties were computed following \cite{Lenz92} for noisy emission line spectra.

\begin{figure}
\includegraphics[width=3.2in]{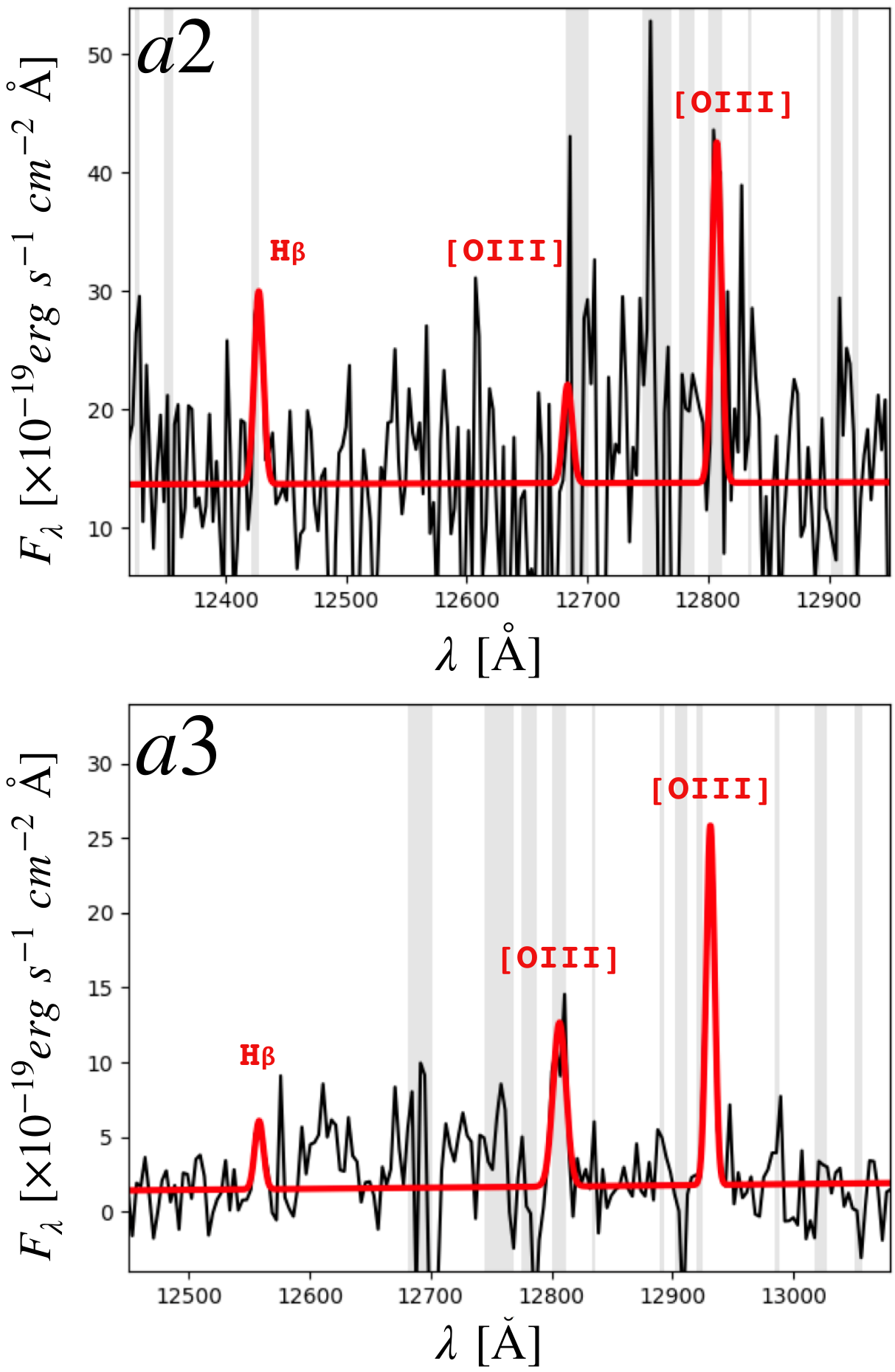}
\caption{Zoom-in of the spectra (smoothed at 70$\rm~km~s^{-1}$) around the H$\beta$ line of the sources \texttt{a2} (J-band SINFONI) and \texttt{a3} (JY-band KMOS): in red the fit to the H$\beta$ and [OIII]$\lambda \lambda$4959,5007\AA~emission lines.}
\label{fig:hboiii}
\end{figure}

For two sources, i.e. \texttt{a2} (one of the BLAGN) and \texttt{a3}, we also detect H$\beta$ and the [OIII]$\lambda 5007$ emission lines in the J-band SINFONI and KMOS spectrum, respectively, as shown in Fig.~\ref{fig:hboiii}. This allowed us to confirm their redshifts and reveal the nature of \texttt{a3} through the BPT diagram \citep{Baldwin81}. 
The estimated integrated flux of the emission lines lead to ratios $\rm log(F_{[OIII]}/F_{H \beta})= 0.7 \pm 0.2$ and $\rm log(F_{[NII]}/F_{H \alpha})= -0.2 _{-0.5} ^{+0.2}$. According to the separation criterion between AGN and SF galaxies in the BPT diagram by \cite{Kauffmann03}, \texttt{a3} is therefore classified as a Type-2 AGN.

Finally, for all 7 galaxies for which a redshift has been measured, we estimated the rest-frame luminosity at 5100\AA~($\rm L_{5100}$) and the V-band absolute magnitudes from the flux measured by interpolating the HST photometric points (i.e., F105W and F140W).
The values, reported in Table~\ref{tab:7emsources}, indicate that such galaxies have high luminosities, i.e. \texttt{a2} and \texttt{b1} are the most powerful sources with $\rm log[L_{5100\AA}/erg~s^{-1}]>44.4$ and $\rm M_V<-23.5$, in agreement with the fact that they host an unobscured AGN. All other sources show slightly lower luminosities, brighter than typical normal galaxies but consistent with what found for galaxies in $z \sim 1.0$ clusters (i.e. close to the knee of the cluster galaxy luminosity function, \citealt{Martinet15}).

\subsection{Dynamics of the galaxy cluster}

\begin{figure}
\includegraphics[width=3.3in]{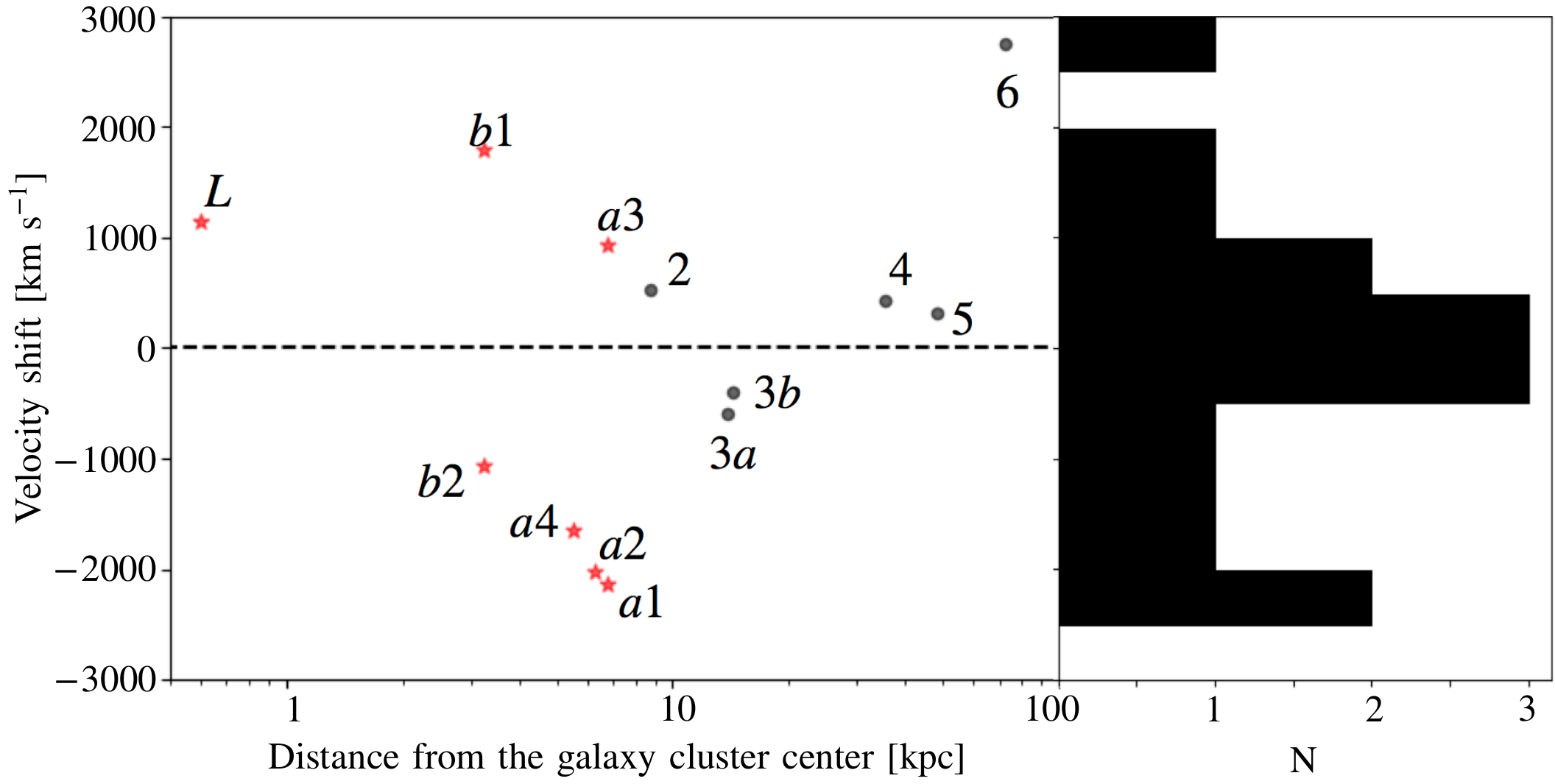}
\caption{Velocity shift of all cluster members as a function of their distance from the X-ray centroid. The histogram of the velocity shifts (binned at 500$\rm~km~s^{-1}$) is reported on the right. Red stars and black dots indicate the cluster members identified in this paper and in \citet{Fassbender14}, respectively.}
\label{fig:VelDist}
\end{figure}

Together with the 7 galaxies belonging to the galaxy cluster discovered in this work, there are additional 6 known cluster members whose redshift has been determined through FORS2 optical slit spectroscopy in \citet{Fassbender14}.
Their ID, coordinates and redshifts are reported in Table~\ref{tab:vel}. Altogether,
XDCP0044 has therefore 13 confirmed cluster members with redshifts in the range 1.5567 < z < 1.5986. The central redshift of the cluster, defined as the mean redshift of all member galaxies, is $z_c =1.5750$ (in agreement with the value $z_c = 1.5790$ published in \cite{Santos11}). For each galaxy member, we estimated the radial component of the peculiar velocity following \citet{Harrison74}:
\begin{equation}
\rm v_{shift} = \frac{(z_i - z_c)}{(1+z_c)} c
\end{equation}
where $1 + z_c$ accounts for the cosmological correction for the cluster Hubble flow and $z_i$ is the redshift of each galaxy belonging to the galaxy cluster (see Table~\ref{tab:vel}). Fig.~\ref{fig:VelDist} reports the velocity-radius diagram (left) and the histogram of the velocity shifts of galaxies (right) binned at 500$\rm~ km~s^{-1}$. We note that, the wide spread of these redshifts $\Delta z \simeq 0.0419$, corresponding to a maximum velocity shift of $\rm \sim 5000~km~s^{-1}$, is due to the presence of few galaxies at the extremes of the distribution. This might be due to the fact that \texttt{XDCP0044} is not completely virialized as already suggested by \cite{Fassbender14}. 
However, the large spread in the redshifts does not result in an extremely large velocity dispersion if compared with literature. 
The cluster velocity dispersion has been estimated in two ways, (1) as the standard deviation and (2) using the statistical gap estimator \citep{Beers90}, which is more robust for small samples (< 20). The derived values are consistent within the errors ($\rm \sigma _c \simeq 1383 \pm 216 ~km~s^{-1}$ and $\rm \sigma _c \simeq 1534 \pm 271 ~km~s^{-1}$, respectively)\footnote{for both methods the uncertainties are computed by adopting the bootstrapping, which is a statistical robust estimator when dealing with small samples (<50).} and comparable with the ones derived for other galaxy clusters at z<1.4 \citep[e.g.][]{Ruel14,Amodeo18}. 
Moreover, the mass of the cluster estimated from its kinematics following \citet{Saro13} ($\rm M_{200} \simeq 4 \times 10^{15}~M_{\odot}$) is higher compared to the one derived through X-ray Chandra data by \cite{Tozzi15} ($\rm M_{200} \simeq 4.4 \times 10^{14} M_{\odot}$). This further suggests that \texttt{XDCP0044} in still not fully virialized and the possible presence of in-falling structures \citep{Bower97,Biviano17}. Additional cluster members will be crucial to fully characterize the velocity dispersion profile.

\begin{table}
\caption{Coordinates, redshift and velocity shift of the SINFONI, KMOS and FORS2 spectroscopic cluster members.} 
\begin{center}                                 
\begin{adjustbox}{width=\columnwidth,center}
\begin{tabular}{c c c c c}         
\hline\hline  
\\[-10pt]
 ID & RA & Dec & z & v\\
\hline
\hline
\multicolumn{5}{c}{SINFONI and KMOS data}  \\
\hline
a1 & 00:44:05.595 & -20:33:53.61 & 1.5567 & -2094 \\
a2 & 00:44:05.607 & -20:33:54.51 & 1.5577 & -1988 \\
a3 & 00:44:05.680 & -20:33:54.92 & 1.5831 &   972  \\
a4 & 00:44:05.590 & -20:33:55.80 & 1.5609 & -1609\\
b1 & 00:44:05.369 & -20:33:56.45 & 1.5904 & 1342   \\
b2 & 00:44:05.388 & -20:33:57.04 & 1.5659 & -1026 \\
L  & 00:44:05.200 & -20:34:00.17 & 1.5848 & 1173 \\
\hline
\hline
\multicolumn{5}{c}{FORS2 data} \\
\hline
2  & 0:44:04.737 & -20:34:09.43 & 1.5795 & 556 \\
3a & 0:44:05.450 & -20:34:16.78 & 1.5699 & -562\\
3b & 0:44:05.531 & -20:34:16.78 & 1.5716 & -364 \\
4  & 0:44:05.325 & -20:33:14.28 & 1.5787 & 463  \\
5  & 0:44:05.611 & -20:32:58.55 & 1.5778 & 358  \\
6  & 0:44:03.015 & -20:32:31.84 & 1.5986 & 2780 \\
\hline
\end{tabular}             
\end{adjustbox}
\end{center}
\label{tab:vel}
\end{table}

\subsection{Chandra X-ray spectroscopy}
\label{sec:xray}

\begin{figure}
\includegraphics[width=0.3\textwidth, angle=270]{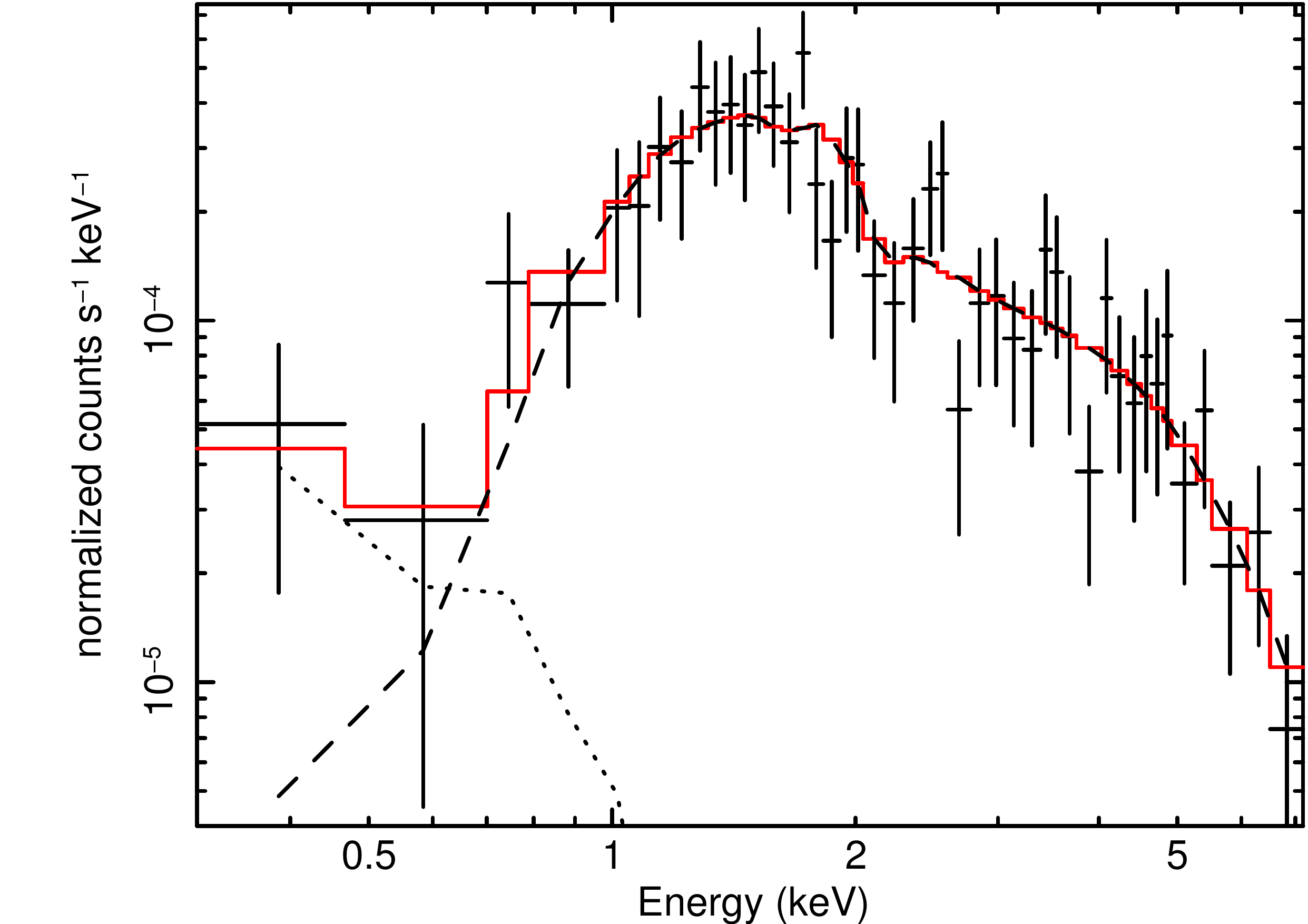}
\centering
\caption{X-ray spectrum and best-fit model (red) of \texttt{b1}. Dotted and dashed lines represent the thermal and absorbed power-law model component, respectively.}
\centering
\label{fig:Xspec}
\end{figure}

Sources \texttt{a2}, \texttt{a3} and \texttt{b1} have been spectroscopically identified as AGN, although only \texttt{b1} has been detected in the Chandra image, as already reported by \citet{Tozzi15}.\\
To measure the nuclear properties of \texttt{b1}, we extracted the X-ray spectrum of this source. The source and local background spectral extractions were performed separately in each observation, using circular regions of 1.5 arcsec radius and annular regions with inner and outer radii of 3 and 7.5 arcsec, respectively\footnote{From this region we excluded a circular region of 1.5 arcsec radius centered on the HST detected position of the BCG galaxy in order to avoid possible, unresolved, contributions from it.}. 
The spectral extractions and response files production were performed with the CIAO script \texttt{specextract}. The spectra were finally co-added using the FTOOLS\footnote{https://heasarc.gsfc.nasa.gov/ftools/} script \texttt{addascaspec}. 
The resulting spectrum has been grouped to 1 count per bin, and modeled in XSPEC v.~12.9.0 in the 0.3-8~keV (0.8-20~keV rest-frame) band  and using the Cash statistics (C-stat) implementation with direct background subtraction \citep{Cash76,Wachter79}. 
The X-ray spectrum of \texttt{b1} consists of 268$\pm21$ (90$\%$ c.l.)  background-subtracted counts, in the 0.3-8 keV energy band and is shown in Fig.~\ref{fig:Xspec}. It has been modeled with an intrinsically absorbed power-law (dashed line) with a thermal emission ({\sc APEC} model in XSPEC, dotted line) at low energies (<0.6 keV) to account for residual excess. The latter component can either parameterize a possible hot thermal corona associated with the quasar host galaxy or may account for the improper background subtraction at these energies due to inhomogeneous distribution of the intracluster-medium in the spectral extraction regions.
The best-fit model ($C-stat/dof=125/173$) gives a $\rm \Gamma=1.4\pm0.2$ and $\rm log[N_{H}/cm^{-2}]=22.7^{+0.1}_{-0.2}$ with a best-fit $\rm kT\approx0.7$~keV (assuming Solar abundance). The latter is unconstrained given that the thermal component affects a restricted spectral region at energies $< 0.6-0.7$~keV of low SNR. The resulting unabsorbed 2-10 keV luminosity is $\rm log [L_{[2-10]keV}/erg~s^{-1}] = 44.1 \pm 0.2$.
A fit with a canonical $\rm \Gamma=1.9$ requires slightly higher column densities of $\rm log[N_{H}/cm^{-2}]=23.0 \pm 0.1$ and results in $\rm log [L_{[2-10]keV}/erg~s^{-1}] = 44.2$. 
\texttt{b1} is, therefore, a luminous and moderately obscured source.

AGN \texttt{a2} and \texttt{a3} have not been detected in Chandra. While this is not at odd for \texttt{a3}, which is classified as a type-2 AGN and its X-ray emission is indeed obscured, \texttt{a2} shows a broad emission line. 
We derived a 3$\sigma$ observed luminosity upper limit at the position of these sources (whose distance is less than Chandra resolution) finding $\rm log[L_{[2-10]keV}/erg~s^{-1}]<43$, assuming an unabsorbed power-law with typical $\Gamma=1.9$.
If we consider the bolometric luminosity of \texttt{a2} derived from $\rm L_{5100 \text{\AA} }$ (i.e. $\rm log [L_{bol}/erg~s^{-1}] \simeq 45.3$; see Sec.~\ref{sec:AGNprop} and Table~\ref{tab:AGNprop}), by applying a bolometric correction $\rm k_{bol,[2-10keV]}\simeq 18.96 \pm 0.24$ \citep{Duras20} we expect an intrinsic luminosity $\rm log[L_{[2-10]keV}/ erg~s^{-1}] \sim 44.01$. 
This means that, to satisfy the upper limit on $\rm log[L_{[2-10]keV}/ erg~s^{-1}]$, a high level of obscuration ($\rm log[N_{H}/cm^{-2}] \gtrsim 23.8$) would be required.
This is consistent with its optical properties. Indeed, while the source \texttt{b1} has an H$\alpha$ emission line clearly broad ($\rm FWHM_{H\alpha}>2000~km/s$), \texttt{a2} has a FWHM which is close to the typical value used as threshold to distinguish between Type-1 and Type-2 AGN ($\rm FWHM_{H\alpha} \sim 1900~km/s$).
The same calculation cannot be made for the \texttt{a3} source, since its host galaxy emission is expected to dominate the 5100\AA\ continuum luminosity and therefore, no bolometric correction can be applied.
The derived X-ray properties are reported in Table~\ref{tab:AGNprop}.

\section{AGN and SF activity in the cluster core} \label{sec:AGNanalysis}

\subsection{Bolometric luminosity, BH masses and Eddington ratios}\label{sec:AGNprop}

Bolometric luminosities were computed by applying the bolometric correction by \citet{Runnoe12} to the 5100\AA\ luminosity, which is estimated from the linear interpolation of the F105W and F140W HST magnitudes given in Table~\ref{tab:FluxHST}. In particular, 
\begin{equation}
    \rm log(L_{bol}) = \zeta \times \lambda L_{\lambda} ~~~~~~~~~~~ \zeta =8.10 \pm 0.42 
\end{equation}
with $\lambda = 5100$\AA.\\
The values obtained for the bolometric luminosity are of the order of $ 10^{45}-10^{46}$ $\rm erg~s^{-1}$ (see Table~\ref{tab:AGNprop}). 
For the source \texttt{b1}, we could also estimate the bolometric luminosity from the [2-10]keV band by assuming the bolometric correction from \cite{Duras20}. The derived value is log[L$\rm_{bol}/ erg~s^{-1}$]=45.4, very close to the one computed from the 5100\AA\ luminosity (i.e. log[L$\rm_{bol}/ erg~s^{-1}$]=45.6).
This agreement confirms that the 5100~\AA\ luminosity is AGN dominated and, therefore, can be used in the BH mass computation. 
Indeed, from the broad H$\alpha$ line, we estimated the BH mass of the two BLAGN (\texttt{a2} and \texttt{b1}) using the virial formula by \citet{GreeneHo}.
\begin{equation}
\rm M_{BH} = 2.0 _{-0.3} ^{+0.4} \times 10^{6}~  \Biggl(  \frac{L_{\rm H\alpha}}{10^{42}} \Biggr) ^{0.55 \pm 0.02} \Biggl( \frac{FWHM}{1000~ \rm{km~s^{-1}}} \Biggr) ^{2.06 \pm 0.06} M_{\odot}
\end{equation}
with $\rm L_{\rm H\alpha}$ defined as in \cite{GreeneHo} to be:
\begin{equation}
\rm L_{\rm H\alpha} = (5.25 \pm 0.02) \times 10^{42} \Biggl( \frac{L_{5100 \AA}}{10^{44} ~\rm{erg~s^{-1}}} \Biggr) ^{1.157 \pm 0.005}
\end{equation}
We found that both \texttt{a2} and \texttt{b1} host relatively massive SMBHs of $\rm \sim 3-8\times 10^7 M_{\odot}$. 
We assign to each BH mass measurement an error given by the sum of the statistical and systematic uncertainties. The systematic uncertainty in the $\rm log(M_{BH})$ determination has been estimated in 0.3 dex to account for the observed scatter in the virial relation itself, while in the computation of the statistical errors, we take into account the errors in the 5100~\AA\ luminosity and FWHM measurements (in quadrature).
From these values we also derived their Eddington ratios, finding $\rm \lambda _{Edd} = L_{bol}/L_{Edd}\sim  0.4-0.5$. \\
Sources \texttt{a2} and \texttt{b1} are therefore luminous, massive, highly accreting and obscured AGN. 
We caution the reader that the values obtained for the BH masses can be underestimated due to the high $\rm N_H$, which might extinguish the H$\alpha$ emission line and the $\rm L_{5100}$. Bolometric luminosities, black hole masses and Eddington ratios are reported in Table \ref{tab:AGNprop}.

\begin{table}
\caption{Physical properties derived for the identified BLAGN.} 
\begin{center}                                  
\begin{adjustbox}{width=\columnwidth,center}
\begin{tabular}{c c c c c c c}          
\hline\hline   
\\[-10pt]
& $ \rm logL_{bol} ^{\star}$ & $ \rm logM_{BH}$ & $\rm \lambda _{Edd}$  &$\rm logL_{X}^{(*)}$ & $\rm logN_{H}$ &$\Gamma$\\[3pt]
&$\rm [erg\,s^{-1}]$&$\rm [M_{\odot}]$&&$\rm [erg\,s^{-1}]$&$\rm [cm^{-2}]$&\\
\hline
a2  & $45.4 \pm 0.2$ & $7.55 _{-0.38} ^{+0.35}$ & $0.51 _{-0.35} ^{+0.68}$ & $44.0$ & $\gtrsim 23.8$&1.9\\ [3pt] 
b1  & $45.6 \pm 0.2$ & $7.86 _{-0.37}^{+0.34}$ &  $0.46_{-0.30} ^{+0.57}$ &44.1 & 22.7&$1.4\pm0.2$\\ [3pt]
\hline
\end{tabular}          
\end{adjustbox}
\end{center}
\label{tab:AGNprop}
$^{\star}$ derived from $L_{5100\AA|}$ following \citet{Runnoe12}; 
$^{(*)}$ the values of $\rm logL_{[2-10]keV}$ reported are intrinsic (unabsorbed). This value has been computed from the X-ray spectrum for \texttt{b1}  while for \texttt{a2} has been derived from the bolometric luminosity. 
\end{table}

\subsection{Star formation activity}  
\label{sec:sfr}
\begin{figure}
\includegraphics[width=3.35in]{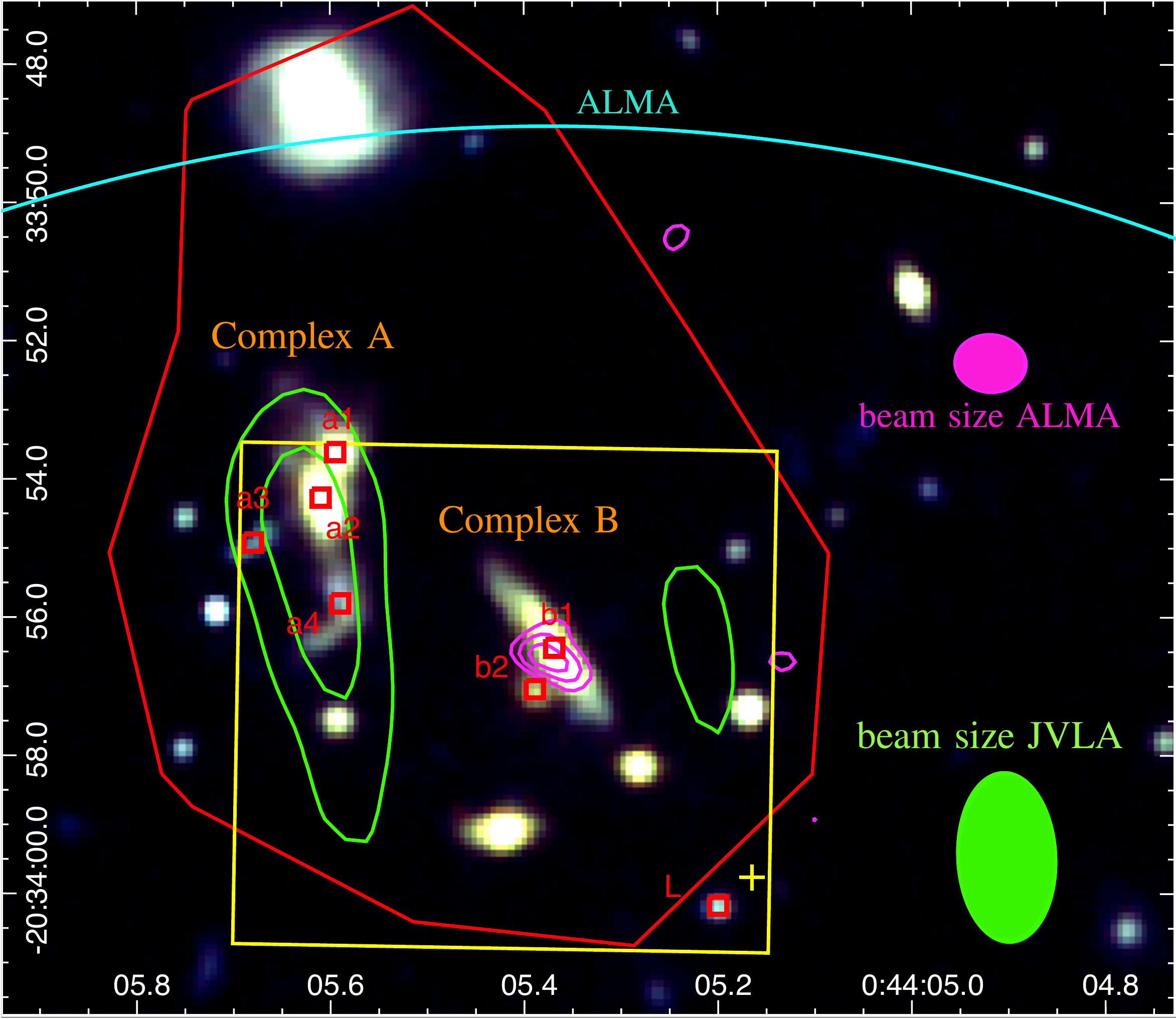}
\caption{HST RGB image of the \texttt{XDCP0044} core, with over-plotted the 3$\sigma$ 1--2 GHz JVLA radio contours (green), the Herschel FIR 3$\sigma$ detection (red) and the 230~GHz ALMA continuum at 5$\sigma$ (magenta). The cyan line indicates the ALMA FOV, while the green and magenta ellipses are the JVLA and ALMA beam, respectively.}
\label{fig:radioima2}
\end{figure}

\citet{Santos15} derived the SFR in the cluster core galaxies through Herschel far-IR data (red contour in Fig.~\ref{fig:radioima2}), finding a SFR $\rm \simeq 450 \pm 60 ~M_{\odot} ~yr^{-1}$. Unfortunately, given the SPIRE point spread function ranging from 17.6$''$ at 250$~\mu$m to 35.2$''$ at 500$~\mu$m, this value includes all sources studied here plus an additional very bright object to the north of the studied region. On the contrary, ALMA observations provide sufficient angular resolution (0.98$''  \times$ 0.78$''$) to distinguish the single sources within the observed ALMA field shown in cyan in Fig.~\ref{fig:radioima2}. 

The analysis of the ALMA data did not spot emission lines in the region covered by the SINFONI FOV. On the contrary, in the ALMA continuum map observed at 230 GHz with an  rms sensitivity of 0.024 mJy/beam (magenta contours in Fig.~\ref{fig:radioima2}), we detected the source \texttt{b1} at 5$\sigma$ significance with a rest-frame 600 GHz flux of $0.80 \pm 0.15$ mJy. No emission is associated to \texttt{a2}, a 3$\sigma$ upper limit on the flux has been computed to be <0.59 mJy.

For the source \texttt{b1}, we derived the SFR by assuming a typical QSO SED, i.e. Mrk 231 template, and normalising it to the observed ALMA flux. The derived value obtained by integrating from 8-to-1000$\rm ~\mu m$, is $\rm 490~M_{\odot}~yr^{-1}$.
This value is consistent with the SFR  derived from IR Herschel data by \citet{Santos15} for the entire central region, suggesting that most of the IR emission, and therefore of the SF, might be associated to \texttt{b1}. However, if we consider a different modelization for the dust emission and we assume e.g. a simple blackbody template normalized to the ALMA flux with dust temperature of 40 K, the SFR we obtain is lower, i.e. $\sim$150 M$_{\odot}$/yr.  Also for source \texttt{a2} we computed an upper limit on the SFR (1) using the Mrk 231 template, finding SFR<285 $\rm M_{\odot}~yr^{-1}$ and (2) by fitting with a simple blackbody template with different temperatures, finding a best fit with $\rm T= 30~K$ and a SFR< 240 $\rm M_{\odot}~yr^{-1}$.

JVLA data reveal an extended radio emission associated to \texttt{Complex A} with a flux density of $\rm S_{1.5~GHz}=0.22 \pm 0.3~mJy$ (green contours in Fig.~\ref{fig:radioima2}). Its luminosity is $\rm L_{1.5~GHz}=3.6\pm0.5\times10^{24}W/Hz$ and under the assumption that the radio signal is produced by a single source, its power ($\rm log [P_{1.5GHz}/W~Hz^{-1}~sr^{-1}]$=23.45) would be just above the threshold at z$\sim$1.6, introduced by \citet[][$\rm log P_{cross} (z)$=21.7+z]{Magliocchetti14}, to discriminate SF processes from AGN, suggesting therefore a likely AGN powered radio emission. However, since the value is slightly above $\rm P_{cross}(z)$ but below $\rm P_{cross}(z)+0.2$, there is a 20$\%$ to 40$\%$ probability that the radio emission is on the contrary due to SF processes \citep{Magliocchetti18}.
In this case the measured radio luminosity would translate into a $\rm SFR \sim 100~ ~M_{\odot} ~yr^{-1}$ according to the relation by \citet{Brown17}, in agreement with the upper limit estimated from the ALMA data.

We then estimated an upper limit on the CO flux, producing a velocity-integrated map by assuming a FWHM=500$\rm~km~s^{-1}$, typically expected for CO emission lines of AGN \citep[e.g.][]{Brusa18,Herrera19}, centred at the redshift of the cluster. This translates into a 3$\sigma$ upper limit on the CO integrated flux for a point source of $\sim$0.27 Jy/beam km/s, equivalent to $\sim 0.585 \rm ~mJy$.
This value corresponds to a $\rm L'_{CO}=4.9 \times 10^{11} K~km/s~pc^2$ following the relation from \citet{Solomon05} and assuming the ratio CO(5-4) to CO(1-0) given by \citet{Carilli13}. According to the values found by \citet{Accurso17} for massive galaxies ($\rm M_{\star} > 10^{11} M_{\odot}$), this value results in $\rm M(CO)<3.5-4 \times 10^9 M_{\odot}$, in agreement with what recently found for galaxies in high-z clusters \citep[e.g.][]{Pulido18,Hayashi18}.

\section{Discussion} \label{sec:discussion}

\subsection{Multiple AGN activity in the cluster core}

As a very interesting result, we found that three out of the 7 galaxies in the core of XDCP0044 host an obscured AGN and that such galaxies are located at a minimum (maximum) projected distance of $\sim$10 kpc (40 kpc). This evidence seems to suggest a crucial role of AGN in the galaxy evolution at this epoch and for these environments. The discovery of multiple AGN activity is common in proto-clusters, where large centrally over-density of AGN are frequently found \citep{Digby10,Galametz10,Krishnan17,Casasola18}, but up to now only few works reported multiple AGN activity in high-z galaxy cluster cores, although not at such small distances.
For example, \citet{Hilton10} reported the detection of two central X-ray cluster member AGN at a projected distance of $\approx$90 kpc in the z = 1.46 galaxy cluster XMMXCS J2215.9-1738 and, at higher redshift (z$\sim$2), \citet{Gobat13} found two X-ray AGN at a projected distance of $\sim$100 kpc in CL J1449+0856.
The newly discovered AGN in \texttt{XDCP0044} are among the closest AGN found at z$>$1 \citep[][and references therein]{DeRosa19} and their proximity implies a future merger between them.
In addition, these AGN are located in merger systems and exhibit high luminosity, high/intermediate obscuration and high accretion rates. Specifically, two of them show broad lines and are mildly ($\rm log(N_H/cm^2) = 22.7$) or highly ($\rm log(N_H/cm^2) > 23.8$) X-ray obscured, while the third one is also optically obscured (type-2 AGN).

Semi-analytical models suggest that mergers between gas-rich galaxies can destabilize the gas and cause its inflow towards the inner regions, thus providing potential accreting material for the central BH and triggering starburst activity \citep[e.g.,][]{Granato04, Hopkins08}. 
This scenario is confirmed by several statistical works \citep[e.g.][]{Alberts16, Martini13} who observed a correlation between AGN and merger activity in both proto-clusters and galaxy clusters.
Moreover, according to this scenario, in this phase the AGN is expected to be highly accreting and obscured \citep[e.g.][]{Koss18}, in agreement with our findings.  

For the two discovered AGN showing broad lines, we also studied their location in the $\rm M_{BH}$ - $\rm M_{*}$ plane. Stellar masses of the AGN host galaxies were computed, as detailed in Sect.~\ref{sec:BCGformation}, finding $\rm log (M_{\star}) \approx 11.3~M_{\odot}$ and $\rm 11.6~M_{\odot}$ for \texttt{a2} and \texttt{b1}, respectively.
As shown in \citet{Bongiorno19}, we found that both of them lie below the \cite{Kormendy13} relation for local inactive galaxies, i.e. at $\sim$2$\sigma$ from it with $\Delta$log($\rm M_{BH}/M_{\star}$) computed perpendicular to the relation. On the contrary, the $\rm M_{BH}/M_{\star}$ ratios seem to be more in agreement with the recent local scaling relation computed for active galaxies at z<0.055 by \cite{Reines15} and with the unbiased $\rm M_{BH}$ - $\rm M_{\star}$ relation derived by \cite{Shankar16}.

\subsection{Mass assembly and time scales for the BCG formation}\label{sec:BCGformation}

How the BCG form is still an open question due to the difficulty in identifying the BCG progenitors at z>1.5. To test the scenario predicting the BCG formation at this epoch through mergers \citep{Stott08,Lidman13,Webb15b,Laporte13}, we estimated the timescale for merging all galaxies discovered to be in the cluster core and the final mass of the possible newly formed galaxy.
We first computed the stellar masses of each galaxy resolved by HAWK-I images \citep[see][]{Fassbender14} by performing an SED-fitting procedure on the HST photometry, HAWK-I J- and K$\rm s$-bands. We used the \textit{zphot} code \citep{Fontana00}, with \citet{Bruzual03} templates, \citet{Salpeter59} IMF and \citet{Calzetti00} extinction.
We adopted exponentially declining star formation histories ($\tau$-models) and approximated the redshift of all the sources to $z_{c}=1.578$. 
Assuming that the 7 confirmed cluster members (all 16 galaxies detected by HST) will be forming the BCG, their best-fit masses would sum up to a final stellar mass of $\rm \sim 1.0\, (2.3) \times 10^{12} M_{\odot}$, consistent with the mass range observed for local BCGs \citep{Zhao15}. 

The BCG assembly time scale is computed according to the average time for a major merger of close pairs reported in \cite{Kitzbichler08}.
We considered the case in which \texttt{Complex A} and \texttt{B} are cluster's sub-clumps, each of which will aggregate to form a cD-like galaxy through a gravitational phase transition. 
In \texttt{Complex B}, this process seems to be already in an advanced state, with \texttt{b1} dominating the system, while in \texttt{Complex A}, several galaxies with similar mass are still visible. However, the resulting merging time ($\rm t_{merge}$) is similar for both complexes, i.e. of the order of $\sim$1-1.5 Gyr. 
This process will lead to the formation of two massive galaxies (galaxy \texttt{A} from \texttt{Complex A} and galaxy \texttt{B} from \texttt{Complex B}) at the center of the galaxy cluster which will possibly merge to form the final BCG. 
We then assumed that these two galaxies and the \texttt{L} source will move towards the X-ray centroid in a dynamical friction time \citep{Binney87} of the order of $ \rm \sim$1.2-2.5 Gyr. 
Therefore, according to this scenario, all these galaxies will merge in $\rm \simeq$ 2.5 Gyr.

A different scenario has also been considered in which \texttt{Complex A} and \texttt{Complex B} do not represent subgroups of the cluster core and therefore all galaxies will directly merge close to the X-ray centroid to form the final BCG in a friction time scale.
In this case, we estimated the mass of the single galaxies resolved by HST observations by assuming a constant mass-to-light ratio in the HST F160W-band, which is the closest to the rest-frame K-band considered as a good indicator of the mass with a 1$\sigma$ scatter of about 0.1 dex \citep{Madau98,Bell03}. According to this scenario, which does not imply the formation of sub-groups, the central QSO \texttt{b1} and the L source are found to have shorter merger times ($\rm < 600~Myr$) compared to all other galaxies, whose friction time ranges from 3.2 to 6 Gyr. Galaxy \texttt{a3} needs more than 10 Gyrs to reach the X-ray centroid and therefore won't merge.\\

Summarising, we find that in a time scale of a couple of Gyrs, all galaxies in the core of XDCP0044 will experience several major mergers, forming a massive central galaxy with $\rm M_{\star}\sim 10^{12}M_{\odot}$ at z$\sim$1, in agreement with what predicted by semi-analytic models \citep{DeLucia07} and found observationally \citep{Prieto15,Sawicki20}. The BCG will then keep slowly growing its mass at z$\leq$1 through minor mergers \citep[e.g.][]{Lidman13,Liu15}.

The newly formed BCG will host a central SMBH, whose lower limit to the mass is $\rm M_{BH} > 2 \times 10^8 M_{\odot}$, obtained by summing the masses of the two AGN, \texttt{a2} and \texttt{b1}.
Stellar and BH masses  of the final BCG as well as merging timescales have to be considered as lower limits. In our computation we have indeed conservatively considered only spectroscopically confirmed galaxies. However, additional cluster complexes (e.g. galaxies complex 3(a+b) identified in \citealt{Fassbender14}) might also take part in the formation of the BCG.

\section{Summary and Conclusions} \label{sec:conclusion}

In this paper we have investigated the properties of the galaxy population in the very central region ($\sim$70 $\times$ 70 kpc$^2$) of \texttt{XDCP0044}, one of the most massive galaxy clusters at $z \sim$1.6. We have analyzed high resolution HST images in F105W, F140W and F160W-bands, IFU spectroscopy obtained with SINFONI in J- and H-band, and KMOS in JY- and H-band, together with JVLA  at 1-2GHz and ALMA band 6 observations at 288~GHz.

\noindent
The main results of our analysis are summarized as follows:
   \begin{enumerate}
      \item[i)] High resolution HST F105W, F140W and F160W images reveal the presence of 16 sources in the core of \texttt{XDCP0044}.
      We find that most of such galaxies are grouped, forming two complexes, i.e. \texttt{Complex A} and \texttt{Complex B}. The first one includes the BCG identified by \cite{Fassbender14} through HAWK-I images, while the second contains the central X-ray AGN identified by \citet{Tozzi15}. 
      
      \item[ii)] Through SINFONI and KMOS spectroscopy, we have confirmed 7 cluster members with redshifts ranging from z=1.5567 to z=1.5904 ($\Delta z \simeq$0.0337), consistently with the redshift of the cluster. 
      In particular, we find that \texttt{Complex A} consists of at least 4 cluster members at a projected distance of $\sim$20 kpc, while the central AGN \texttt{b1}, and the nearby ($\sim$5 kpc) galaxy \texttt{b2}, belong to the \texttt{Complex B}.

      \item[iii)] In 2 of the 7 confirmed cluster members (\texttt{a2} in \texttt{Complex A} and \texttt{b1} in \texttt{Complex B}) we detect a broad ($\rm \gtrsim 2000~km~ s^{-1}$) H$\alpha$ emission line. These sources have been therefore classified as BL-AGN, hosting massive ($\rm > 10^7 - 10^8 M_{\odot}$) and highly accreting ($\rm \sim 0.4-0.5$) BHs. 
Moreover, the analysis of the BPT diagram pointed out the presence of an additional AGN which does not show broad lines (a3 in \texttt{Complex A}). The minimum distance in this AGN triple is 10 kpc between a2 and a3, while the maximum distance between a3 and b1 is 40 kpc.
  
      \item[iv)] One of the BLAGN, i.e. \texttt{b1}, was already identified as AGN from its unresolved X-ray emission by \citet{Tozzi15}. The analysis of the spectrum reveals that  \texttt{b1} is a luminous and moderately obscured AGN, with  X-ray luminosity $\rm L_{[2-10]keV}= 1.2_{-0.6} ^{+1.4}\times 10^{44}\rm  erg\,s^{-1}$ and column density $\rm N_H \simeq 5.4 \times 10^{22} cm ^{-2}$. The other two objects (the BLAGN \texttt{a2} and the NLAGN a3), on the contrary, are not detected in the Chandra data, implying a high level of obscuration. In particular, source \texttt{a2}, for which the intrinsic X-ray luminosity could be derived from the bolometric one, is found to be X-ray luminous ($\rm  L_{[2-10]keV} \sim 10^{44} ~erg~s^{-1}$) and obscured ($\rm log [N_{\rm H}/cm^{-2}] \gtrsim 23.8$).
      
      \item[v)]  The integrated SFR of the whole central region of \texttt{XDCP0044} is $\rm \simeq 450 \pm 60 ~M_{\odot} ~yr^{-1}$. This value has been derived by \citet{Santos15} using Herschel data which however do not allow us to distinguish the different sources. Thanks to the higher resolution ALMA observations, we derive the SFR of the single observed sources finding that the central AGN \texttt{b1} alone contributes to this value with a SFR ranging from 150 to 490 $\rm M_{\odot}/yr$, depending on the assumed SED, while \texttt{a2} might contribute less than $\sim$285 $\rm M_{\odot}$/yr, in agreement with the radio emission of \texttt{Complex A}.

\end{enumerate}

In conclusion, \texttt{XDCP0044} allows us to witness the BCG assembly in one of the densest galaxy cluster core at z$\sim1.6$, which is thought to be in a crucial formation epoch, when both SF and nuclear activity are at their peak \citep{Madau14,Aird15}. We confirm that high-z galaxy cluster cores show different properties compared to the z=0 ones. Indeed, no single, early-type BCG has been detected in the core of \texttt{XDCP0044}, which is found to host a large number (at least 7 confirmed) of highly star-forming interacting galaxies grouped in two main complexes, both hosting multiple AGN activity. The discovered AGN triple is one of the closest revealed so far at z>1 \citep{DeRosa19}, with a  projected distance ranging from 10 to 40 kpc.
Moreover, these results lead to a scenario in which obscured AGN activity is triggered during the formation of the cluster BCG, when mergers between gas-rich galaxies provide the fuel for the AGN and for triggering starburst activity. According to our data, we expect to form a typical massive galaxy of $\rm M_{\star}\sim 10^{12} M_{\odot}$, hosting a SMBH with mass $> 2 \times 10^8 ~M_{\odot}$, in a time scale of few Gyrs.\\

\section*{Acknowledgements}

We thank Andrea Biviano, Claudio Ricci and Alessandra Lamastra for useful discussions. This work is based on observations collected at the European Southern Observatory under ESO programmes 094.A-0713(A) and 092.A-0114(A).  AB, EP, LZ and MB acknowledge the support from ASI-INAF 2017-14-H.0. PT acknowledges support from the Istituto Nazionale di Astrofisica (INAF) PRIN-SKA 2017 program 1.05.01.88.04 (ESKAPE).
This paper makes use of the following ALMA data: ADS/JAO.ALMA$\#$2017.1.01387.S. ALMA is a partnership of ESO (representing its member states), NSF (USA) and NINS (Japan), together with NRC (Canada), MOST and ASIAA (Taiwan), and KASI (Republic of Korea), in cooperation with the Republic of Chile. The Joint ALMA Observatory is operated by ESO, AUI/NRAO and NAOJ.

\section*{Data availability}
All data are free in the archive of the different telescopes, while data generated in this work are available on request.

\bibliographystyle{mnras}
\bibliography{andrea} 

\appendix

\section{SINFONI and KMOS spectra of cluster members} \label{appendix:spectra}

Zoom-in of the SINFONI J and H-band  and KMOS JY- and H-band (for \texttt{a3}) rest-frame spectra of the galaxies in the field of interest classified as cluster members.

\begin{figure*}
\centering
\includegraphics[width=6in]{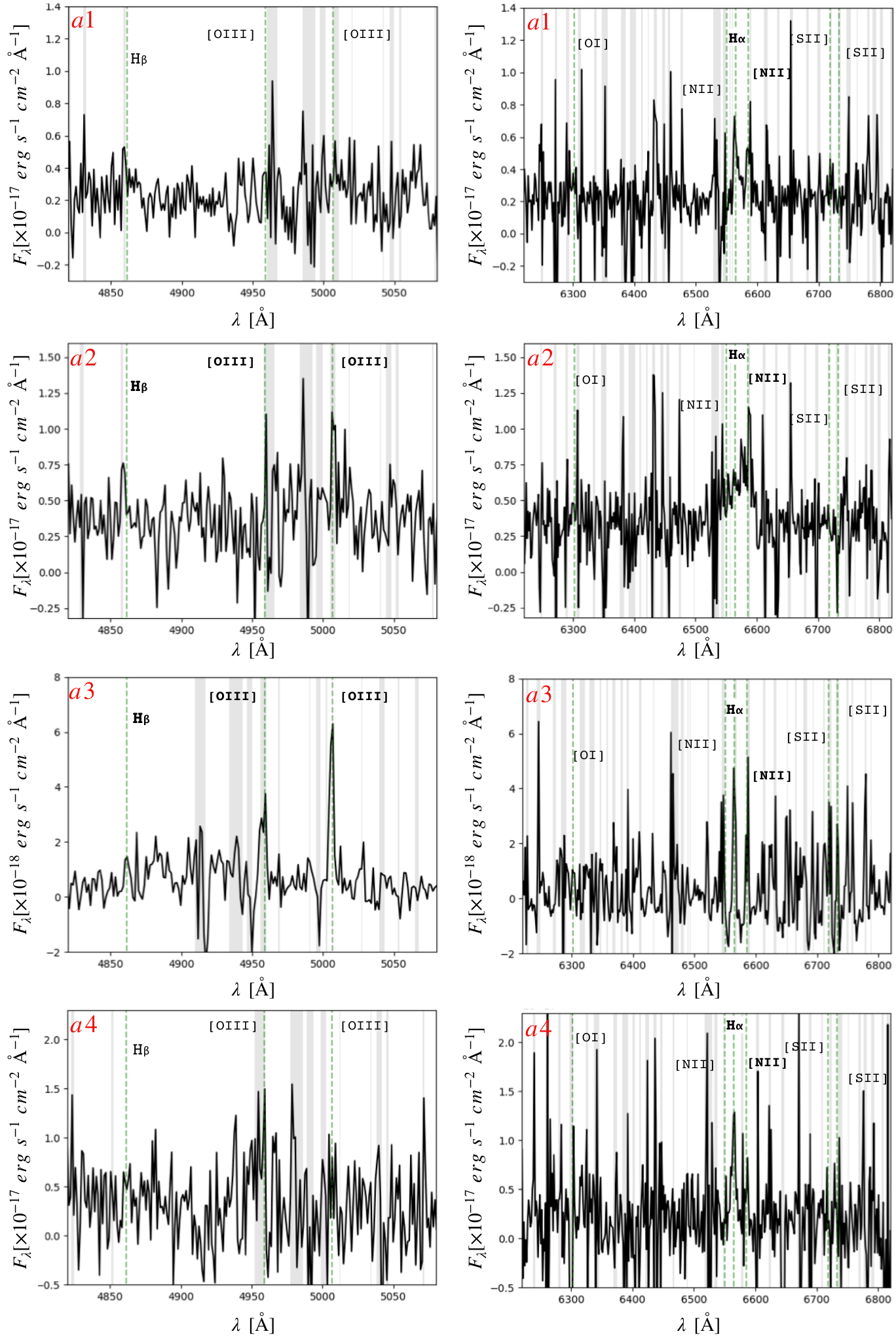}
\caption{Zoom-in of the SINFONI J and H-band (KMOS JY- and H-band for \texttt{a3}) rest-frame spectra of the galaxies in \texttt{Complex A} in spectral regions in which (left panels) [OIII] doublet and H$\beta$ emission lines and (right panels) [NII] and [SII] doublet, [OI] and H$\alpha$ emission lines, are expected to be observed. The expected position for each emission line is marked with a dashed green line and bold labels correspond to the detected lines. The grey vertical bands indicate the wavelength regions contaminated by the sky emission lines. All the spectra are binned at 70~$\rm km~s^{-1}$.}
\label{fig:spectraTot1}
\end{figure*}

\begin{figure*}
\centering
\includegraphics[width=6in]{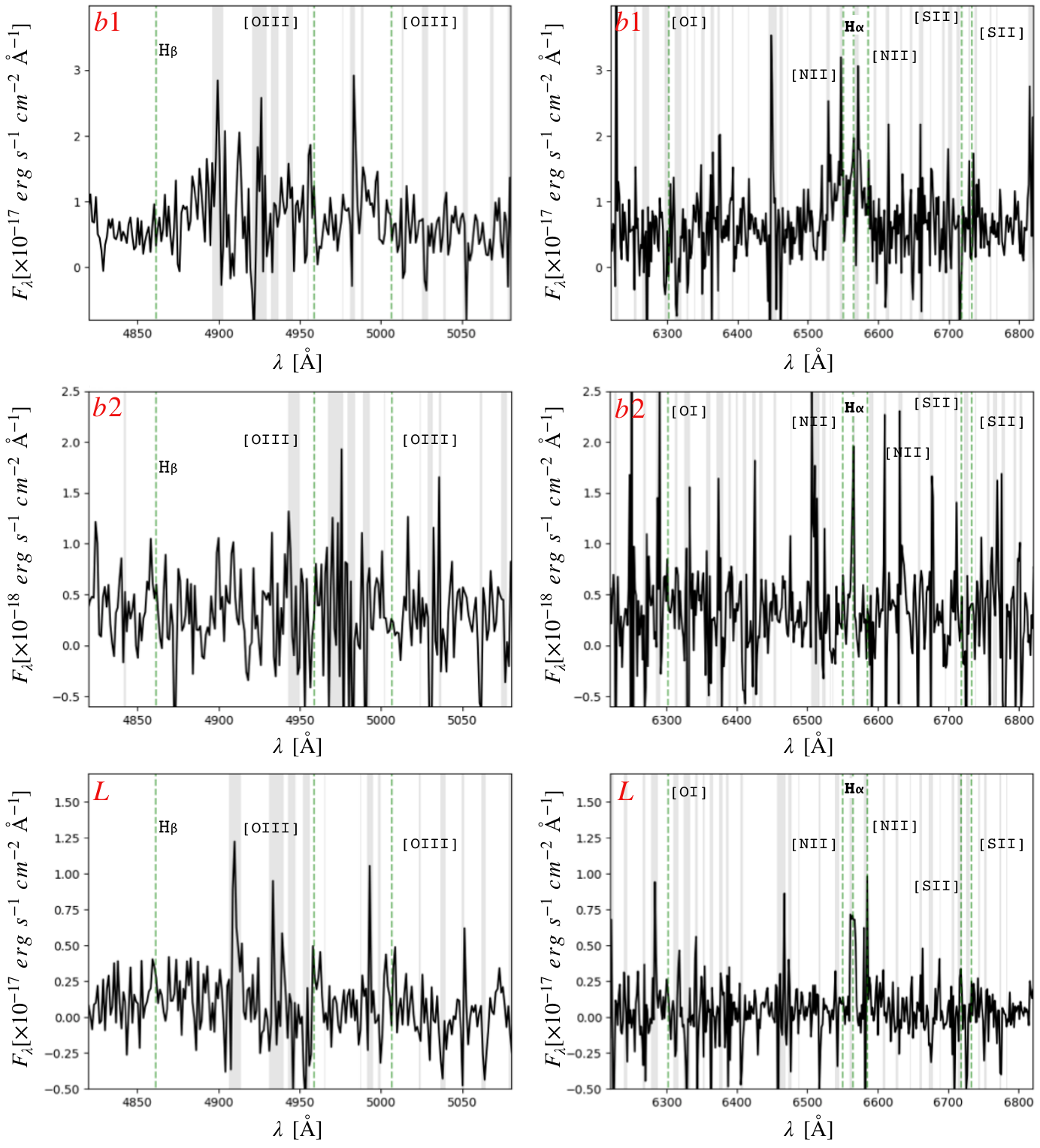}
\caption{Same as Fig.~\ref{fig:spectraTot1} but for \texttt{Complex B} and \texttt{L} source.}
\label{fig:spectraTot2}
\end{figure*}

\section{JVLA radio detected sources in XDCP0044}
\label{sec:radio}

\begin{figure*}
\centering
\includegraphics[width=5.5in]{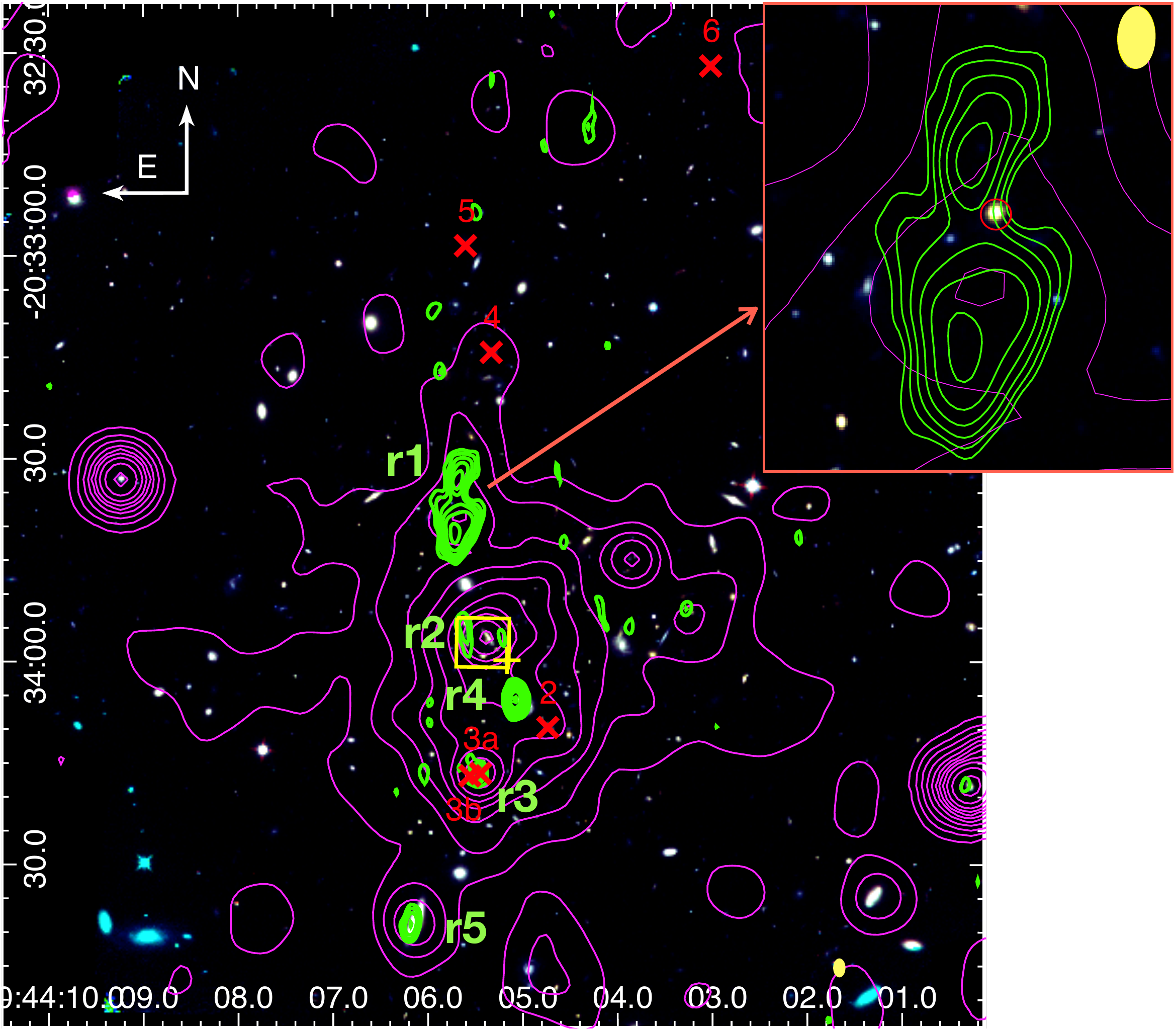}
\caption{HST RGB image of XDCP0044 with $\rm 3\sigma$ 1--2 GHz JVLA radio emissions contours (green) and soft X-ray emission (magenta). Yellow filled ellipse corresponds to  the beam size of the JVLA data. The red crosses indicate the confirmed cluster members from FORS2 spectroscopy by \citet{Fassbender14}. Five radio sources are indicated with r1, r2, r3, r4, and r5. The top-right panel shows a zoom-in of the radio source r1 and its most probable optical counterpart (marked with a red circle), which was identified by \citet{Fassbender14} as a passive galaxy. The yellow ellipse represents the beam size of the JVLA observations.}
\label{fig:radioima}
\end{figure*}

Radio JVLA data include the whole galaxy cluster field. Here, we briefly summarize the radio sources detected in XDCP0044.
Fig.~\ref{fig:radioima} shows the HST RGB image of the cluster with overlapped the 3$\sigma$ radio JVLA contours in green at 1.5 GHz and the soft X-ray emission in magenta.
In addition to the extended radio emission associated with \texttt{Complex A} and described in Sec.\ref{sec:sfr} (i.e. r2), four radio sources have been detected, i.e. three compact radio sources in the south region of the galaxy cluster (\texttt{r3}, \texttt{r4} and \texttt{r5}) and an extended emission, \texttt{r1}, in the north part. 
We report the JVLA fluxes of the identified radio sources in Table \ref{tab:radio}. 

The compact radio sources \texttt{r5} and \texttt{r4} are associated with optically detected galaxies for which the redshift is unknown. However, the optical counterpart of \texttt{r5} is probably a foreground galaxy, since its size at z$\geq$1.5 would be too large ($\sim$35 - 50 kpc) compared to a typical galaxy.
On the contrary, the compact radio emission \texttt{r3} is associated to a group of interacting galaxies hosting an X-ray AGN found to be part of the cluster \citep[object 3 in][]{Tozzi15}. Its radio luminosity is $\rm L_{1.5~GHz}= 2.6\pm 0.3 \times 10^{24}~W~Hz^{-1}$, not k-corrected.\\

Finally, the hourglass-like extended emission , \texttt{r1}, does not show a radio core or a central radio galaxy. It might be associated to one or multiple optical counterparts located in the central narrowing, none of which  has a known redshift. The most probable candidate seems  the passive galaxy identified by \citealt{Fassbender14} (see zoom-in of Fig.~\ref{fig:radioima}).
This radio source is spatially correlated to an X-ray north-south elongation reported by \citet{Tozzi15} and described as evidence of mass accretion onto the cluster. This claim is supported by the presence of three spectroscopically confirmed cluster members (red crosses in figure) located along the north axis \citep{Fassbender14}.

We extracted the radio spectral index\footnote{Defined as $S_{\nu}\propto \nu^{\alpha}$.} $\alpha$ of this source combining JVLA data at 1.5 GHz with a detection from the TGSS at 150 MHz \citep{Intema17}. We found $\alpha = -1.3 \pm 0.1$, steeper than the average spectral index associated with radio galaxy lobes. This might be interpreted as a signature of aged plasma from a remnant radio galaxy.
Assuming this radio source at the distance of the cluster, its radio luminosity at 1.5 GHz would be $7.0\pm0.2\times10^{25}$~W/Hz, k-corrected assuming the measured $\alpha$. 
This luminosity is high if compared with typical radio galaxies in local clusters \citep{vanVelzen12}. However, to date we know very little about the typical luminosities of high-redshift radio galaxies in cluster environments.
Alternatively, this source might be directly associated to the Intra-Cluster medium (ICM), possibly generated by compression of old plasma bubbles from merging induced shock waves (the so-called "radio phoenixes"; \citealt{Ensslin01}). This would explain the steep spectrum of the source. However, the high luminosity is at odds with what expected for these kind of sources \citep{VanWeeren19}.

\begin{table}
\centering
\caption{JVLA 1-2 GHz fluxes of the identified radio sources.}           
\centering                                      
\begin{tabular}{c c}          
\hline\hline  
\\[-10pt]
$\rm Radio\, source$ & $ \rm S_{[1.5~GHz]}$ \\
  & $\rm [mJy]$ \\
\hline
r1 & 3.19$\pm$0.07 \\
r2 & 0.22$\pm$0.03 \\
r3 & 0.16$\pm$0.02 \\
r4 & 0.85$\pm$0.03 \\
r5 & 0.28$\pm$0.03 \\
\hline
\end{tabular} \\[3pt]
\label{tab:radio}
\end{table}

\bsp	


\label{lastpage}
\end{document}